\documentclass[preprint,fleqn,showpacs,showkeys]{revtex4}
\usepackage{graphicx}
\usepackage{amssymb}
\usepackage{amsmath}
\usepackage{bm}
\usepackage{feynmf}

\def\be{\begin{equation}}
\def\ee{\end{equation}}
\def\bea{\begin{eqnarray}}
\def\eea{\end{eqnarray}}
\def\wh{\widehat}

\begin{document}
\setcounter{page}{1}
\title{Interpolating Scattering Amplitudes between Instant Form and Front Form of 
Relativistic Dynamics}
\author{Chueng-Ryong \surname{Ji}}
\author{Alfredo Takashi \surname{Suzuki}}
\thanks{Present address: Southern Adventist University, Collegedale, TN - 37315.\\${}^{*}$Permanent address: Instituto de F\'{\i}sica Te\'orica-UNESP Universidade Estadual Paulista,\\
Rua Dr. Bento Teobaldo Ferraz, 271 - Bloco II - 01140-070, S\~ao Paulo, SP, Brazil.}
\affiliation{Department of Physics, North Carolina State University, Raleigh, NC 27695-8202}
\author{ \surname{}}
\affiliation{}

\date[]{}

\begin{abstract}
Among the three forms of relativistic Hamiltonian dynamics proposed by Dirac in 1949, the instant form 
and the front form can be interpolated by introducing an interpolation angle between the ordinary time 
$t$ and the light-front time $(t+z/c)/\sqrt{2}$. Using this method, we introduce the interpolating scattering amplitude
that links the corresponding time-ordered amplitudes between the two forms of dynamics
and provide the physical meaning of the kinematic transformations
as they allow the invariance of each individual time-ordered amplitude for an arbitrary interpolation angle. 
In particular, it exhibits that the longitudinal boost is kinematical only in
the front form dynamics, or the light-front dynamics (LFD), but not in any other interpolation angle dynamics.
It also shows that the disappearance of the connected contributions to the current arising from the vacuum occurs
when the interpolation angle is taken to yield the LFD. Since it doesn't require the infinite momentum 
frame (IMF) to show this disappearance and the proof is independent of reference frames, it resolves the confusion between the LFD and the IMF.  The well-known utility of IMF usually discussed in the instant form dynamics 
is now also extended to any other interpolation angle dynamics using our interpolating scattering amplitudes.
\end{abstract}



\maketitle

\section{Introduction}
\label{sec.01}


When the particle systems have the characteristic momenta which are of the same order or even much larger than the
masses of the particles involved, it is part of nature that a relativistic treatment is called for in order to describe those systems properly.
In particular, relativistic effects are most essential to describe the low-lying hadron systems in terms of
strongly interacting quarks/antiquarks and gluons in quantum chromodynamics (QCD).
For the study of relativistic particle systems, Dirac proposed the three different forms of the relativistic Hamiltonian dynamics
in 1949 \cite{Dirac}: i.e. 
the instant ($x^0 =0$), front ($x^+ = (x^0 + x^3)/\sqrt{2} = 0$), and point ($x_\mu x^\mu = a^2 > 0, x^0 > 0$) forms.
While the instant form dynamics (IFD) of quantum field theories is based on the usual equal time $t=x^0$ quantization ($c=1$ unit is taken here), 
the equal light-front time $\tau \equiv (t + z/c)/\sqrt{2}=x^+$ quantization yields the front form dynamics, more commonly 
called light-front dynamics (LFD), correspondingly. Although the point form dynamics has also been 
explored \cite{Glozman}, the most popular choices were thus far the equal-$t$ (instant form) and 
equal-$\tau$ (front form) quantizations.

A crucial difference between the instant form and the front form
may be attributed to their energy-momentum dispersion relations. When a particle has
the mass $m$ and the four-momentum $k = (k^0, k^1, k^2, k^3)$, the relativistic energy-momentum
dispersion relation of the particle at equal-$t$ is given by
\begin{equation}
\label{eq.01}
k^0 = \sqrt{{\vec k}^2 + m^2}, 
\end{equation}
where the energy $k^0$ is conjugate to $t$ and the three-momentum vector ${\vec k}$ is given by ${\vec k} =
(k^1, k^2, k^3)$. However, the corresponding energy-momentum dispersion relation at equal-$\tau$ is
given by
\begin{equation}
\label{eq.02}
k^- = \frac{{\vec k}^2_\perp + m^2}{k^+} , 
\end{equation}
where the light-front energy $k^-$ conjugate to $\tau$ is given by $k^- = (k^0 - k^3)/\sqrt{2}$ and the light-front
momenta $k^+ = (k^0 + k^3)/\sqrt{2}$ and ${\vec k}_\perp = (k^1, k^2)$ are orthogonal to $k^-$. 
While the instant form (Eq.(\ref{eq.01})) exhibits an irrational
energy-momentum relation, the front form (Eq.(\ref{eq.02})) yields a rational relation and
thus the signs of $k^+$ and $k^-$ are correlated, e.g. the momentum $k^+$ is always positive when
the system evolve to the future direction (i.e. positive $\tau$ ) so that the light-front energy
$k^-$ is positive. In the instant form, however, no sign correlations for $k^0$ and ${\vec k}$ exist. Such
a difference in the energy-momentum dispersion relation makes the LFD quite distinct from other forms of the relativistic Hamiltonian dynamics.

The light-front quantization \cite{Dirac,Steinhardt} has already been 
applied successfully in the context of current algebra \cite{lcca}
and the parton model \cite{lcparton} in the past.
With further advances in the Hamiltonian renormalization 
program \cite{brodsky,wilson1}, LFD appears to be 
even more promising for the relativistic treatment of hadrons.
In the work of Brodsky, Hiller and McCartor \cite{Hil},
it is demonstrated how to solve the problem of renormalizing light-front
Hamiltonian theories while maintaining Lorentz symmetry and other
symmetries. The genesis of the work presented in \cite{Hil} may be found in
\cite{RM} and additional examples including the use of LFD methods to
solve the bound-state problems in field theory can be found in the 
review of  QCD and other field theories on the light cone \cite{BPP}. 
These results are indicative of the great 
potential of LFD 
for a fundamental description of non-perturbative effects in strong 
interactions. This approach may also provide a bridge between the two
different pictures of hadronic matter, i.e. the
constituent quark model (CQM) (or the quark parton model) closely
related to experimental observations and the QCD based on a covariant non-abelian quantum field theory.
Again, the key to possible connection between the two pictures is the
rational energy-momentum dispersion relation given by Eq.(\ref{eq.02})
that leads to a relatively simple vacuum structure. There is no 
spontaneous creation of massive fermions in the LF quantized vacuum. 
Thus, one can immediately obtain a constituent-type 
picture \cite{GJC} in which all 
partons in a hadronic state are connected directly to the hadron instead 
of being simply disconnected excitations (or vacuum fluctuations) in a 
complicated medium. A possible realization of chiral symmetry breaking in 
the LF vacuum has also been discussed in the literature \cite{Wilson}.

Moreover, the Poincar{\'e} algebra in the ordinary equal-$t$ 
quantization is drastically changed in the light-front equal-$\tau$ 
quantization. In LFD, the maximum number (seven) of the ten Poincare
generators are kinematic (i.e. interaction independent) and they leave
the state at $\tau = 0$ unchanged~\cite{5}. However,
the transverse rotation whose direction is perpendicular to the
direction of the quantization axis $z$ at equal $\tau$ becomes a dynamical problem 
in LFD because the quantization surface $\tau$ is not invariant
under the transverse rotation and  
the transverse angular momentum operator involves the interaction
that changes the particle number \cite{surya}. 
Leutwyler and Stern showed that the angular momentum operators can be 
redefined to satisfy the SU(2) spin algebra and the commutation relation 
between mass operator and spin operators~\cite{6};
\begin{equation}
[{\cal{J}}_i,{\cal{J}}_j] = i {\epsilon_{ijk}} {\cal{J}}_k,
\end{equation}
\begin{equation}
[M,{\cal{\vec J}}] =0.
\end{equation}
Nonetheless, in LFD, there are two dynamic equations to solve:
\begin{equation}
{\cal J}^2 |H; p^+, {\vec p_{\perp}}^2> =
{S_H}({S_H}+1) |H; p^+, {\vec p_{\perp}}^2>
\end{equation}
and
\begin{equation}
M^2 |H; p^+, {\vec p_{\perp}}^2> =
{m_H}^2 |H; p^+, {\vec p_{\perp}}^2>,
\end{equation}
where the total angular momentum (or spin) and the
mass eigenvalues of the hadron ($H$) are given by
$S_H$ and $m_H$.
Thus, it is not a trivial matter to specify
the total angular momentum of a specific hadron state.

As a step towards understanding the conversion of the dynamical problem
from boost to rotation, we constructed the Poincar{\'e} 
algebra interpolating between instant and light-front time quantizations~\cite{chad}.
We used an orthogonal coordinate system which interpolates smoothly 
between the equal-time and the light-front quantization hypersurface. 
Thus, our interpolating coordinate system had a nice feature of tracing 
the fate of the Poincare algebra at equal time as the hypersurface 
approaches to the light-front limit.
The same method of interpolating hypersurfaces has been used by Hornbostel
\cite{Hornbostel} to analyze various aspects of field theories including the issue of nontrivial vacuum.
The same vein of application to study the axial anomaly in the Schwinger model has also
been presented \cite{JiRey}, and other related works \cite{Chen, Elizalde, Frishman, Sawicki} can also be found in the literature.

In the present work, we introduce the interpolating scattering amplitude
that links the corresponding time-ordered amplitudes between the two forms of dynamics.
We exemplify the physical meaning of the kinematic 
transformations in contrast to the dynamic transformations by means of checking the invariance of each individual time-ordered
amplitude for an arbitrary interpolation angle. Our analysis further clarifies why and how 
the longitudinal boost is kinematical only in the LFD but not in any other interpolation angle dynamics including IFD. 
In particular, we show the disappearance of the connected contributions to the current arising from the vacuum 
when the interpolation angle is taken to yield the LFD. Since we don't need any infinite momentum 
frame (IMF) to show this disappearance and our proof is completely independent of reference frames,
it resolves the confusion between the LFD and the IMF that sometimes appears
in the discussion on related topics. The well-known utility of IMF usually discussed in the instant form dynamics is now also extended to any 
other interpolation angle dynamics using our interpolating scattering amplitudes.

In the next section, Section \ref{sec.02},  we introduce the interpolating scattering amplitude
that links the corresponding time-ordered amplitudes between the two forms of dynamics and
show the disappearance of the connected contributions to the current arising from the vacuum 
when the interpolation angle is taken to yield the LFD. Taking just the simplest possible example (viz. spin-less scalar particles)
and keeping only the fundamental degrees of freedom, i.e. particle momenta, we focus only on the essential part of the time-ordered
scattering amplitude, namely the energy denominators. In Section \ref{sec.03}, we discuss the kinematic transformations that allow the invariance of each individual time-ordered amplitude for an arbitrary interpolation angle and present the explicit results of particle momenta 
under those kinematic transformations. In this section, we also discuss a remarkable difference of the LFD result compared to the result for any other interpolation angle dynamics including IFD and the role of the longitudinal boost that becomes kinematic only in LFD. 
In Section \ref{sec.04}, we explicitly show the invariance of the individual time-ordered amplitude under kinematic transformations plotting the results obtained in Section 
\ref{sec.03} and extend the well-known utility of IMF in IFD to an arbitrary interpolation angle dynamics.  Conclusions follow in Section\ref{sec.05}.

\section{Interpolating Scattering Amplitudes}
\label{sec.02}

We begin by adopting the following convention of the space-time coordinates to define the interpolating angle:
\bea
\label{interpolangle}
\left[ \begin{array}{c}
x^{\widehat{+}}  \\
x^{\widehat{-}}
\end{array}
\right] = \left[ \begin{array}{cc}\cos\delta & \sin\delta \\
\sin\delta & -\cos\delta \end{array} \right]\left[\begin{array}{c} x^0 \\ x^3  \end{array}\right],
\eea 
and 
\bea
\left[ \begin{array}{c}
x^0  \\
x^3
\end{array}
\right] = \left[ \begin{array}{cc}\cos\delta & \sin\delta \\
\sin\delta & -\cos\delta \end{array} \right]\left[\begin{array}{c} x^{\wh+} \\ x^{\wh-}  \end{array}\right],
\eea 
in which the interpolating angle is allowed to run from 0 through $45^\circ$, $0\le \delta \le \frac{\pi}{4}$. 
All the indices with the wide-hat notation signify the variables with the interpolation angle $\delta$.
For the limit $\delta \rightarrow 0$ we have $x^{\wh+} = x^0$ and $x^{\wh-} = -x^3$ so that we recover 
usual space-time coordinates although the z-axis is inverted while for the other extreme limit, $\delta \rightarrow \frac{\pi}{4}$ we have $x^{\wh{\pm}} = (x^0\pm x^3)/\sqrt{2} \equiv x^{\pm}$ which 
leads to the standard light-front coordinates.  
Of course, the same interpolation applies to the momentum variables:
\bea
\label{interpolangle-momentum}
\left[ \begin{array}{c}
p^{\widehat{+}}  \\
p^{\widehat{-}}
\end{array}
\right] = \left[ \begin{array}{cc}\cos\delta & \sin\delta \\
\sin\delta & -\cos\delta \end{array} \right]\left[\begin{array}{c} p^0 \\ p^3  \end{array}\right].
\eea 
For any two interpolating four vector variables $a^{\wh{\mu}}$ and $b^{\wh{\mu}}$, the scalar product $a_{\wh{\mu}} b^{\wh{\mu}}$ must be identical to $a_\mu b^\mu$ and is given by
\be
\label{scalar-product}
a_{\wh{\mu}} b^{\wh{\mu}} = (a^{\wh+}b^{\wh+}-a^{\wh-}b^{\wh-})\cos2\delta+(a^{\wh+}b^{\wh-}+a^{\wh-}b^{\wh+})\sin2\delta-a^{\wh1}b^{\wh1}-a^{\wh2}b^{\wh2}.
\ee
We may define
\bea
{\mathbb C} & = & \cos2\delta, \\ \nonumber
{\mathbb S} & = & \sin2\delta, \\ \nonumber
{\vec {\bf a}}_{\wh{\perp}} & = & a^{\wh1} \hat {\bf{x}}+ a^{\wh2} \hat {\bf {y}},
\eea
for shorthand notations and convenience, so that the Minkowski space-time metric $g^{\wh{\mu}\wh{\nu}} = (\wh+,\,\wh-,\,\wh1,\,\wh2)$ with interpolating angle may be written as  
\bea
g^{\wh{\mu}\wh{\nu}} =
\left[ \begin{array}{rrrr}
\mathbb{C} & \mathbb{S} & 0 & 0  \\
\mathbb{S} & -\mathbb{C} & 0 & 0 \\
0 & 0 & -1 & 0 \\
0 & 0 & 0 & -1 
\end{array} \right] = g_{\wh{\mu}\wh{\nu}}.
\eea
Thus, the covariant and contravariant indices are related by
\bea
\label{definitions}
a_{\wh+} =  \mathbb{C}a^{\wh+}+\mathbb{S}a^{\wh-} & ; & a^{\wh+}  =  \mathbb{C}a_{\wh+}+\mathbb{S}a_{\wh-} \\ \nonumber
a_{\wh-} =  \mathbb{S}a^{\wh+}-\mathbb{C}a^{\wh-} & ; & a^{\wh-}  =  \mathbb{S}a_{\wh+}-\mathbb{C}a_{\wh-}\\ \nonumber
a_{\wh{j}} = -a^{\wh{j}} & , &  \mbox{($j=1,\,2$)}.
\eea
As the coordinate variable $x^+$ plays the role of the time evolution parameter and the canonical conjugate energy variable is $p_+=p^-$ 
in LFD, we also take $x^{\wh +}$ to be the evolution parameter and the conjugate energy variables with the corresponding subscripts, e.g., 
$q_{\wh + }$.

\begin{figure}[tt]
\begin{center}
\includegraphics[scale=.8]{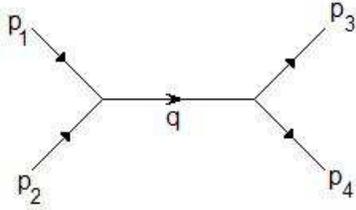}
\caption{Scattering amplitude of spinless particles.}
\label{fig:1} 
\end{center}      
\end{figure}

Now, we discuss the scattering amplitude of two spin-less particles, e.g. an analogue of the well known QED annihilation/production process $e^+ e^- \to \mu^+ \mu^-$ in a toy $\phi^3$ model theory, as depicted in Fig.\ref{fig:1}. In this work we do not involve spins and any other degrees of freedom except the fundamental degrees of freedom, i.e. particle momenta, for the simplest possible illustration.

Although we discuss here just this simple scattering amplitude, the bare-bone structure that we demonstrate in this analysis will be commonly applicable to any extended calculation of amplitudes including other degrees of freedom. In particular, not only the basic structure of the amplitudes but also the fundamental degrees of freedom to describe the scattering process will prevail in such extension. Further complications from other degrees of freedom beyond the particle momenta would appear separately without modifying the energy denominator structure that we discuss in this work: 
e.g., the terms associated with the spin degrees of freedom in QED would appear as the matrix elements in the numerator but not in the denominator of the amplitude. The extension of the present work to the gauge field theories involving other degrees of freedom such as QED and QCD is in progress. In this work, we will focus on the basic structure of the scattering amplitudes, i.e. the energy denominators, considering only the fundamental degrees of freedom, i.e. particle momenta.
       
\begin{figure}[tt]
\begin{center}
\includegraphics[scale=.8]{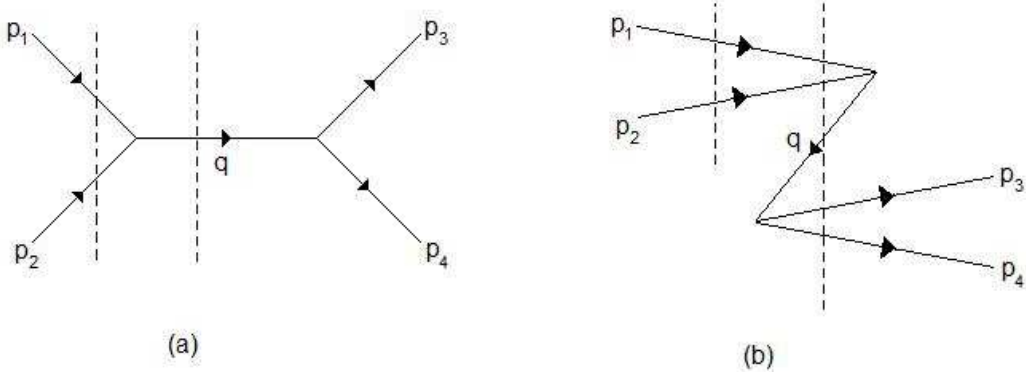}
\caption{Time-ordered amplitudes in IFD for the Feynman amplitude depicted in Fig.\ref{fig:1}.}
\label{fig:2} 
\end{center}      
\end{figure}

Modulo inessential factors including the  square of the coupling constant, the lowest order tree-level Feynman diagram 
shown in Fig.\ref{fig:1} is proportional to the propagator of the intermediate particle, that is,
\be
\Sigma =  \frac{1}{s-m^2}
\ee
where $s = (p_1+p_2)^2$ is the Mandelstam variable which is invariant under any Poincar{\'e} transformations and $m$ is the mass of the intermediate boson.  
Of course, the physical process can take place only above the threshold $s > 4 M^2$, where $M$ is the mass of 
the final particle and anti-particle that are produced, e.g. like the muon mass in the $e^+e^- \rightarrow \mu^+ \mu^-$ scatterring process. 
In the IFD, where the initial conditions are set on the hyperplane $t=0$ and the system evolves with the ordinary time  $t>0$, this manifestly covariant Feynman amplitude is decomposed into the corresponding two time-ordered amplitudes, graphically represented in Fig.\ref{fig:2}(a) and Fig.\ref{fig:2}(b).
%
These two time-ordered amplitudes correspond respectively to the following analytic expressions:
\be
\Sigma^{a}_{\rm IFD} = \frac{1}{2q^0} \left(\frac {1}{p_1^{0}+p_2^0-q^0} \right ),
\ee
and
\be
\Sigma^{b}_{\rm IFD} = -\frac{1}{2q^0} \left(\frac {1}{p_1^{0}+p_2^0+q^0} \right ).
\ee
It is not difficulty to show that the sum of the time-ordered amplitudes is identical to the manifestly covariant Feynman amplitude:
\bea
\label{IFD_amp}
\Sigma_{\rm IFD} & = & \Sigma_a^{\rm IFD} + \Sigma_b^{\rm IFD}  \nonumber \\
                 & = & \frac{1}{2q^0} \left(\frac {1}{p_1^{0}+p_2^0-q^0} -\frac {1}{p_1^{0}+p_2^0+q^0}\right ) \nonumber \\
                 & = & \frac{1}{s-m^2},
\eea
where the conservation of the three momentum ${\vec p}_1+{\vec p}_2 = {\vec q}$ as well as the energy-momentum dispersion relation $q^0 = \sqrt{{\vec q}^2 + m^2}$ in IFD is used to get the covariant denominator $s -m^2$ in the last step.

To obtain the corresponding time-ordered amplitudes in an arbitrary interpolating angle $\delta$, we just need to change 
the superscript 0 of the IFD energy variables in the energy denominators to the superscipt ${\wh +}$ and multiply an overall factor $\mathbb C$ to the amplitudes: i.e.
\be
\label{sigma_interpol_a}
\Sigma^{a}_{\delta} = \frac{1}{2q^{\widehat +}} \left(\frac{{\mathbb C}}{p_1^{\widehat +}+p_2^{\widehat +}-q^{\widehat+}}\right ),
\ee
and
\be
\label{sigma_interpol_b}
\Sigma^{b}_{\delta} = -\frac{1}{2q^{\widehat +}} \left(\frac {{\mathbb C}}{p_1^{\widehat +}+p_2^{\widehat +}+q^{\widehat+}}\right ).
\ee
The overall factor $\mathbb C$ is necessary because the energy of the particle with the four-momentum $p^{\widehat \mu}$ in an arbitrary interpolation angle is given by $p_{\widehat +}$ while 
the contravariant $p^{\widehat +}$ used in the interpolating amplitudes is related to the covariant $p_{\widehat +}$
with the factor $\mathbb C$ as shown in Eq.(\ref{definitions}), i.e. 
$p^{\widehat +}={\mathbb C}p_{\widehat +}+{\mathbb S}p_{\widehat -}$.
Note here that the factor ${\mathbb S}$ in front of the longitudinal momentum $p_{\widehat -}$ is irrelevant because the longitudinal momenta of the initial particles must be cancelled by the longitudinal momentum of the intermediate particle due to the conservation of the longitudinal momentum.  
Again, it is not so difficulty to show that the sum of the time-ordered amplitudes for any angle $\delta$ is identical to the manifestly covariant Feynman amplitude:
\bea
\label{IAD_amp}
\Sigma_{\delta} & = & \Sigma_{\delta}^a + \Sigma_{\delta}^b  \nonumber \\
                 & = & \frac{1}{2q^{\widehat +}} \left(\frac{{\mathbb C}}{p_1^{\widehat +}+p_2^{\widehat +}-q^{\widehat+}} -\frac {{\mathbb C}}{p_1^{\widehat +}+p_2^{\widehat +}+q^{\widehat+}}\right ) \nonumber \\
                 & = & \frac{1}{s-m^2},
\eea
where we used the relation between the covariant and contravariant indices (see Eq.(\ref{definitions})) such as 
$q^{\widehat +}={\mathbb C}q_{\widehat +}+{\mathbb S}q_{\widehat -}$ and the conservation of momenta $p_{1\wh-}+p_{2\wh-}=q_{\wh-}$ and ${\vec {\bf p}}_{1\wh{\perp}} +{\vec {\bf p}}_{2\wh{\perp}}={\vec {\bf q}}_{\wh{\perp}}$ as well as the four-momentum scalar product relation (see Eq.(\ref{scalar-product})) to get the Lorentz invariant denominator $s -m^2$ in the last step.
It is also rather easy to see that Eq.(\ref{IAD_amp}) becomes Eq.(\ref{IFD_amp}) as ${\mathbb C}$ goes to the unity. 
In LFD however, i.e. as ${\mathbb C}$ goes to zero, the denominator in the first amplitude $\Sigma^a_{\delta=\pi/4}$, i.e. $1/(p_1^{\widehat +}+p_2^{\widehat +}-q^{\widehat+})=1/(p_1^+ + p_2^+ - q^+)$ goes to infinity due to the conservation $p_1^+ + p_2^+ = q^+$ but the multiplication of ${\mathbb C}=0$ with this infinity makes the finite result $1/(s-m^2)$, while the second amplitude $\Sigma^b_{\delta=\pi/4}$ is wiped out
due to ${\mathbb C}=0$. This result is akin to the very well-known result from the work entitled
``Dynamics at Infinite Momentum'' \cite{Weinberg}. 
However, we would like to make it clear that the disappearance of the second amplitude $\Sigma^b_{\delta=\pi/4}$ 
in LFD is different from the usual IMF result obtained by taking $P_z \rightarrow  \pm \infty$ with $P \equiv p_1+p_2$ for a shorthand notation (e.g. 
$P^2 = s$).  As far as any correlation between the interpolation angle $\delta$ and the total longitudinal momentum $P_z$ 
is avoided, our derivation is completely independent of the frame and the 
only relevant parameter to show this disappearance is the interpolation angle $\delta$ which has nothing to do with the choice of
reference frame. In Section \ref{sec.04}, we will discuss the special case with a particular correlation between $\delta$ and $P_z$
and the associated treacherous point similar to the zero-mode issue in LFD. 

For the rest of this section, we elaborate more details of our derivations discussed above. 
The dispersion relation $q^2 = m^2$ in terms of interpolating angle variables results in a quadratic equation in $q_{\wh +}$ and $q_{\wh - }$ that can be solved for the energy variable $q_{\wh +}$ in terms of momentum components $q_{\wh -}$ and $\vec{\bf q}_{\perp}$ as well as mass $m$:
\be
\label{drsolution}
q_{\wh +} = \frac {-\mathbb{S}q_{\wh -}\pm \omega_q }{\mathbb{C}},
\ee
in which we introduced the notation
\be
\label{omegaq}
\omega_q = \sqrt{q_{\wh{-}}^2 + \mathbb{C}\left(\vec{\bf q}_{\wh{\perp}}^2+m^2 \right)}.
\ee
For the physical solution with positive energy in Eq. (\ref{drsolution}), we must take
\be
q_{\wh +} = \frac {-\mathbb{S}q_{\wh -}+ \omega_q }{\mathbb{C}},
\ee
which identifies $\omega_q$ as 
\be
\omega_q = \mathbb{C}q_{\wh +} + \mathbb{S}q_{\wh -} = q^{\wh +}  \, .
\ee
For $\delta = 0$ and $\delta = \frac{\pi}{4}$, $\omega_q $ becomes $q^0 = \sqrt{\vec{\bf q}^2 + m^2}$ and $q^+ = (q^0 + q^3)/\sqrt{2}$ , respectively. Using this variable $\omega_q$, we may rewrite Eqs. (\ref{sigma_interpol_a}) and (\ref{sigma_interpol_b}) as follows:
\bea
\Sigma_{\delta}^a & = & \frac{1}{2\omega_q D_+} \nonumber \\
\Sigma_{\delta}^b & = & \frac{1}{2\omega_q D_-},
\eea
where
\bea
\label{D+D-}
D_+ & = & P_{\wh+} + \frac{\mathbb{S} q_{\wh-} - \omega_q}{\mathbb{C}} \nonumber \\
D_- & = & P_{\wh+} + \frac{\mathbb{S} q_{\wh-} + \omega_q}{\mathbb{C}}\,,
\eea
in which we used the longitudinal momentum conservation $P_{\wh -} = (p_1)_{\wh-}+(p_2)_{\wh -} = q_{\wh -}$. The sum of both contributions given by Eq. (\ref{IAD_amp}) can then be expressed as
\bea
\label{sum}
\Sigma_{\delta} & = & \Sigma^a_{\delta} + \Sigma^b_{\delta} \nonumber \\
                & = & \frac{1}{2\omega_q} \left(\frac {1}{P_{\wh+}+\frac{\mathbb{S}q_{\wh-}-\omega_q}{\mathbb{C}}} - \frac {1}{P_{\wh+}+\frac{\mathbb{S}q_{\wh-}+\omega_q}{\mathbb{C}}} \right) \, ,
\eea
which is identical to the second line of Eq.(\ref{IAD_amp}). In Eq. (\ref{sum}), we can confirm $\Sigma_\delta = 1/(s-m^2)$:
\bea
\Sigma_{\delta} & = & \frac{\frac{1}{\mathbb{C}}}{\left(P_{\wh+}+\frac{\mathbb{S}q_{\wh-}}{\mathbb{C}} \right)^2-\left(\frac{\omega_q}{\mathbb{C}} \right)^2} \nonumber \\
                 & = & \frac{1}{\mathbb{C}P_{\wh+}^2+2\mathbb{S}P_{\wh+}q_{\wh-}+\frac{\mathbb{S}^2q_{\wh-}^2}{\mathbb{C}}-\frac{\omega_q^2}{\mathbb{C}}} \nonumber \\
                 & = & \frac{1}{\mathbb{C}P_{\wh+}^2+2\mathbb{S}P_{\wh+}P_{\wh-}-\mathbb{C}P_{\wh-}^2-\vec{\bf P}_{\wh{\perp}}^2-m^2} \nonumber \\
           & = & \frac{1}{s-m^2} \, ,
\eea
where we used $\omega_q^2 = q_{\wh-}^2+\mathbb{C}(\vec{\bf q}_{\wh{\perp}}^2 + m^2)$, $P_{\wh-} = q_{\wh-}$ and $\vec{\bf P}_{\wh{\perp}} = \vec{\bf q}_{\wh{\perp}}$.
%
Using Eq. (\ref{sum}), we may now recapture the instant form and light-front limits, as follows. 

For the instant form limit (IFD), we have $\delta \rightarrow 0$ (i.e. $\mathbb{C} \rightarrow 1$ and $\mathbb{S} \rightarrow 0$) and $\omega_q \rightarrow q_{\wh+}$. 
In this limit, it is apparent that Eq. (\ref{sum}) becomes
\be
\Sigma_{\delta \rightarrow 0} \equiv \Sigma_{IFD} = \frac{1}{2q_{\wh+}}\left(\frac{1}{P_{\wh+}-q_{\wh+}} - \frac{1}{P_{\wh+}+q_{\wh+}}\right)
=   \frac{1}{2q_{0}}\left(\frac{1}{P_{0}-q_{0}} - \frac{1}{P_{0}+q_{0}}\right) \, ,
\ee
where $\delta = 0$ is taken in the interpolating angle variables.

For the light-front limit (LFD), $\delta \rightarrow \frac{\pi}{4}$ (i.e. $\mathbb{C} \rightarrow 0$ and $\mathbb{S} \rightarrow 1$), we expand
$\omega_q$ given by Eq.(\ref{omegaq}) in the orders of $\mathbb{C}$ and get
\be
\omega_q \rightarrow q_{\wh-} + \frac{\mathbb{C}\left(\vec{\bf q}_{\wh{\perp}}^2+m^2 \right)}{2q_{\wh-}}+ {\cal O}(\mathbb{C}^2).
\ee
Substituting this expansion of $\omega_q$ in the denominator of the first term in Eq. (\ref{sum}), we get
\bea
\frac{\mathbb{S}q_{\wh-}-\omega_q}{\mathbb{C}} & \rightarrow & -\frac{\vec{\bf q}_{\wh{\perp}}^2 + m^2}{2q_{\wh-}}+{\cal O}(\mathbb{C}) \nonumber \\
                                               & \rightarrow & -\frac{\vec{\bf q}_{\wh{\perp}}^2 + m^2}{2q_{\wh-}}\quad \mbox{as $\mathbb{C} \rightarrow$ 0}.
\eea 
For the second denominator in Eq. (\ref{sum}), however, we get
\bea
\frac{\mathbb{S}q_{\wh-}+\omega_q}{\mathbb{C}} & \rightarrow & \frac{2}{\mathbb{C}}-\frac{\vec{\bf q}_{\wh{\perp}}^2 + m^2}{2q_{\wh-}}+{\cal O}(\mathbb{C}) \nonumber \\
                                               & \rightarrow & \infty \quad \mbox{as $\mathbb{C} \rightarrow$ 0}.
\eea 
Thus, in the light-front limit ($\mathbb{C} \rightarrow 0$), the contribution from the second diagram vanishes and
\be
\label{LFD}
\Sigma_{\delta \rightarrow \frac{\pi}{4}} = \frac{1}{2q_{\wh-}}\frac{1}{\left\{P_{\wh+}-\frac{(\vec{\bf q}_{\wh{\perp}}^2+m^2)}{2q_{\wh-}}\right\}}
= \frac{1}{P^{+}}\frac{1}{\left\{P^{-}-\frac{(\vec{\bf P}_{\perp}^2+m^2)}{2P^{+}}\right\}} \, ,
\ee 
where $q_{\wh-} \rightarrow q_{-} = q^+$ and $\vec{\bf q}_{\wh{\perp}} \rightarrow \vec{\bf q}_\perp$ are same with $P^+$ and $\vec{\bf P}_\perp$, respectively, due to the momentum conservation in LFD.
Again, we would like to make it clear that the disappearance of the second amplitude $\Sigma^b_{\delta=\pi/4}$ 
in LFD is different from what has been known from the usual IMF, i.e. $P_z \rightarrow  \pm \infty$.
As we will discuss in the next section, Section \ref{sec.03}, the longitudinal boost is kinematic in LFD so that 
the disappearance of the connected contribution $\Sigma_{\delta \rightarrow \frac{\pi}{4}}^b$ to the current arising from the vacuum 
is independent of $P_z$ or the IMF. This is certainly not the case for any other interpolation case, i.e. $\delta \neq \pi/4$.
The longitudinal boost becomes dynamic for $\delta \neq \pi/4$ and the contributions from $\Sigma^a_\delta$ and $\Sigma^b_\delta$
depend on $P_z$ (or the reference frames) and the well-known utility of IMF can be extended to an arbitrary interpolating angle $0 \le \delta < \frac{\pi}{4}$. We will discuss more on this point in Section \ref{sec.04} after we present the physical meaning of the kinematic
transformations in Section \ref{sec.03}.   
\vspace{1cm}

\section{Kinematic Transformations of Particle Momenta}
\label{sec.03}



As we presented in the previous section, Section \ref{sec.02}, the sum of all the time-ordered amplitudes (just two in our example discussed in Section~\ref{sec.02}) must be independent of
the interpolation angle $\delta$ and identical to the manifestly covariant Feynman amplitude. Although the total amplitude is 
Poincar\'e invariant, the individual time-ordered amplitude is neither invariant in general nor independent of $\delta$. 
Thus, one may ask a question if the individual time-ordered amplitude can be invariant at least under some subset of
Poincar\'e generators. The answer is yes and this issue is what we would like to address in this section.
The point is that the individual time-ordered amplitude would not change as far as the time evolution parameter  
$x^{\widehat +}$ doesn't change so that the individual time-ordered amplitude would be invariant under a certain
transformation which doesn't alter the time evolution parameter $x^{\widehat +}$. To the extent that the time evolution parameter $x^{\widehat +}$ doesn't change, all the momentum components with ${\widehat +}$ such as $q^{\widehat +}$
would not change because the same transformation rules apply to both the space-time coordinates and the four-momenta
of the particles involved. Such subset of the Poincar\'e group that doesn't alter the time evolution parameter $x^{\widehat +}$ is known as the stability group. Since the transformations that belong to the stability group do not modify the time evolution parameter $x^{\widehat +}$, each time-ordered amplitude must be invariant under these transformations. 
Individual time-ordered amplitudes represent the dynamics given at each instant of time defined by the time evolution parameter $x^{\widehat +}$ in the given form of the relativistic quantum field theory.
For this reason, it may be appropriate for the transformations that leave each individual time-ordered amplitudes 
invariant to be called as the {\em kinematic transformations} and 
the generators of those transformations belong to the stability group deserve to be distinguished 
from the other Poincar\'e group generators. All other Poincar\'e group generators besides the kinematic generators
are dynamical and change the contributions from each individual time-ordered amplitudes. 
In this section, we discuss the kinematic transformations for an arbitrary interpolation angle $\delta$.  
In particular, we take the limits to $\delta = 0$ and $\pi/4$ to discuss the fates of the kinematic transformations
in the two distinguished forms of the relativistic dynamics, IFD and LFD, respectively.
Since we focus mainly on the fundamental dynamic variables not involving any other degrees of freedom 
(e.g. spins) in this work, our results of the kinematic transformations apply explicitly only to the particle momenta.
 
The matrix of the homogeneous part of Poincar\'e group in the interpolating angle basis may be written\cite{chad} as
\bea
\label{matrix}
M_{\wh{\mu}\wh{\nu}} = 
\left[ \begin{array}{cccc}
0 & K^3 & {\mathcal D}^{\wh1} & {\mathcal D}^{\wh2}  \\
-K^3 & 0 & {\mathcal K}^{\wh1} & {\mathcal K}^{\wh2} \\
-{\mathcal D}^{\wh1} & -{\mathcal K}^{\wh1} & 0 & J^3 \\
-{\mathcal D}^{\wh 2}  & -{\mathcal K}^{\wh 2} & -J^3 & 0 
\end{array} \right]
\eea
where 
\bea
\label{generators3}
{\mathcal K}^{\wh1}  =  -K^1\sin\delta-J^2\cos\delta & ; & {\mathcal K}^{\wh 2}  =  J^1\cos\delta-K^2\sin\delta  \nonumber \\
{\mathcal D}^{\wh1}  =  -K^1\cos\delta+J^2\sin\delta & ; & {\mathcal D}^{\wh2}  =  -J^1\sin\delta-K^2\cos\delta \, .  
\eea
The kinematic generators ${\mathcal K}^{\wh j}$ and the dynamic ones ${\mathcal D}^{\wh j}, \, j=(1,\,2)$, can also be written as the combinations of $E^{\wh j}$ and $F^{\wh j}$:
\bea
\label{generators1}
{\mathcal K}^{\wh 1}  =  {\mathbb C}F^{\wh 1} - {\mathbb S}E^{\wh 1} & ; &  {\mathcal K}^{\wh 2} =  {\mathbb C}F^{\wh 2} - {\mathbb S}E^{\wh 2} \nonumber \\
{\mathcal D}^{\wh 1}  =  -{\mathbb S}F^{\wh 1} - {\mathbb C}E^{\wh 1} & ; &  {\mathcal D}^{\wh 2}  =  -{\mathbb S}F^{\wh 2} - {\mathbb C}E^{\wh 2}\, ,
\eea
where
\bea
\label{generators2}
E^{\wh 1}  =  J^2\sin\delta+K^1\cos\delta & ; & E^{\wh 2}  =  K^2\cos\delta-J^1\sin\delta \nonumber\\
F^{\wh 1}  =  K^1\sin\delta-J^2\cos\delta & ; & F^{\wh 2}  =  J^1\cos\delta+K^2\sin\delta \, . 
\eea
The interpolating operators $E^{\wh j}$ and $F^{\wh j}$ coincide with the usual $E^{j}$ and $F^{j}$ of LFD
in the limit $\delta = \pi/4$.  
As discussed in Ref.\cite{chad}, the transverse boosts ($K^1, K^2$) are dynamic whereas the transverse rotations ($J^1, J^2$) are kinematic in IFD ($\delta = 0$),
while the LF transverse boosts ($E^1, E^2$) are kinematic whereas the LF transverse rotations ($F^1, F^2$)
are dynamic in LFD ($\delta = \frac{\pi}{4}$).  
One may note the swap of the roles between ``boosts'' and ``rotations'' in the two forms of relativistic dynamics, IFD and LFD,
and utilize it for some hadron phenomenology\cite{carl}.

We may check explicitly that the generators ${\mathcal K}^{\wh j}$ given above satisfy the commutation relation 
$\left[ {\mathcal K}^{\wh{j}}, {\mathcal P}^{\wh+} \right] = 0$ with the momentum operator ${\mathcal P}^{\wh +}$
using Eq. (\ref{generators1}) and the interpolating Poincar\'e algebra presented in Ref.\cite{chad} :
\bea
\label{commutatorzero}
\left[{\mathcal K}^{\wh{j}}, \, {\mathcal P}^{\wh{+}}\right] & = & {\mathbb C}\left[F^{\wh {j}}, \, {\mathcal P}^{\wh{+}}\right]-{\mathbb S}\left[E^{\wh{j}},\,{\mathcal P}^{\wh{+}}\right] \nonumber\\
& = & {\mathbb C}\left(-i{\mathcal P}^{\wh{j}}{\mathbb S}\right)-{\mathbb S}\left(-i{\mathcal P}^{\wh{j}}{\mathbb C}\right) = 0.
\eea
This means that each transformation of the form $\exp{(-i\omega{\mathcal K}^{\wh{j}})}$, ($j=1,\,2$), leaves the momentum operator ${\mathcal P}^{\wh+}$ invariant. As a consequence if the momentum $P^{\wh+}$ is an eigenvalue of the operator ${\mathcal P}^{\wh +}$, $P^{\wh+}$ remains invariant under the cited transformations. Likewise, the plus ($\wh+$) component of any four vector is invariant under such transformations and the time $x^{\wh+}$ remains invariant as well.  It verifies that the generators ${\mathcal K}^{\wh{j}}$ are kinematic. 

In a similar way, for the generators ${\mathcal D}^{\wh{j}}$, we may check explicitly that  the commutators $\left[ {\mathcal D}^{\wh{j}},{\mathcal P}^{\wh+} \right]$ are now nonvanishing:
\bea
\label{commutatornonzero}
\left[{\mathcal D}^{\wh{j}}, \, {\mathcal P}^{\wh{+}}\right] & = & -{\mathbb S}\left[F^{\wh {j}}, \, {\mathcal P}^{\wh{+}}\right]-{\mathbb C}\left[E^{\wh{j}},\,{\mathcal P}^{\wh{+}}\right] \nonumber\\
& = & -{\mathbb S}\left(-i{\mathcal P}^{\wh{j}}{\mathbb S}\right)-{\mathbb C}\left(-i{\mathcal P}^{\wh{j}}{\mathbb C}\right) = i{\mathcal P}^{\wh{j}}.
\eea
Since commutators above are not only nonvanishing but also proportional to ${\mathcal P}^{\wh {j}}$, each transformation of the form $\exp{(-i\omega{\mathcal D}^{\wh{j}})}$, ($j=1,\,2$), develops 
transverse components of the momentum and cannot 
leave the momentum $P^{\wh+}$ invariant.
Thus, the generators ${\mathcal D}^{\wh{j}}$ are dynamic.

Among the elements involved in the matrix given by Eq.(\ref{matrix}), it is interesting to note that the rotation
around the longitudinal direction, i.e. $J^3$, is unique because it doesn't change $x^{\wh +}$ and thus kinematic
for any interpolation angle $\delta$. However, the longitudinal boost $K^3$ has a quite different characteristic
compared to any other operators in Eq.(\ref{matrix}).    
To see this, let's look at the commutator between ${\mathcal P}^{\wh +}$ and $K^3$ in the Poincar\'e algebra:
\bea
\left[{\mathcal P}^{\wh+},K^3\right] & = & i{\mathcal P}_{\wh-} \nonumber \\
          & = & i\left( \mathbb{S}{\mathcal P}^{\wh+}-\mathbb{C}{\mathcal P}^{\wh-} \right ) \, ,
\eea  
which leads to 
\bea
\label{K3P+}
\left[{\mathcal P}^{+},K^3\right] & = & i {\mathcal P}^{+} 
\eea  
in the limit $\delta \rightarrow \frac{\pi}{4}$. This shows that the longitudinal boost has a distinguished property
in the limit $\delta \rightarrow \frac{\pi}{4}$, namely it becomes kinematic in this limit. 
Although the right hand side of Eq.(\ref{K3P+}) doesn't vanish, it yields the same ${\mathcal P}^+$ operator in the commutation
relation. This means that the eigenvalues of ${\mathcal P}^+$ operator, or the LF longitudinal momentum $P^+$, are just scaled
by the factor $e^{\beta_3}$ when it is boosted in the longitudinal direction by the rapidity $\beta_3$. 
By the same token, the LF energy $P^-$ is scaled by the factor $e^{-\beta_3}$ under the same transformation
due to the commutation relation in LFD,
\bea
\label{K3P-}
\left[{\mathcal P}^{-},K^3\right] & = & - i {\mathcal P}^{-} \, . 
\eea  
It may be interesting to note that the algebra among ${\mathcal P}^+, {\mathcal P}^-$ and $K^3$ works just the similar way as
the algebra among the creation, annihilation and number operators in one-dimensional simple harmonic oscillator.
Due to the conservation of three momenta ($P^+ , {\vec P}_\perp$) as well as the compensating scale factors of
$e^{-\beta_3}$ and $e^{\beta_3}$ between the LF energy ($P^-$) and the LF longitudinal momentum ($P^+$),
one can show that each individual LF time-ordered amplitudes are invariant under the longitudinal boost $K^3$.
This may be also understood from the intactness of the LF time $x^+$ modulo the same scaling factor $e^{\beta_3}$
for the LF longitudinal momentum under the $K^3$ operation. With this reasoning, one may understand that
$K^3$ becomes the kinematic generator in LFD although it is dynamical for any other interpolation angle
$0 \le \delta < \pi/4$. As the boost problem in IFD is one of the most difficult problems to deal with in the relativistic
many-body calculations, all of the boost operators ($K^1, K^2, K^3$) have been known as difficult operators
in IFD.  Since at least $K^3$ can change its difficult characteristic to a good one, i.e. from dynamic to kinematic,
and joins the stability group in LFD, one may regard such dramatic character change of $K^3$ in LFD as a kind of
``return of a prodigal son''. Of course, the community of LFD welcomes the addition of $K^3$ in the stability group.  
For this reason, the number of kinematic generators in LFD is one more than all other cases of 
interpolating angles in the range $0 \le \delta < \frac{\pi}{4}$ as shown in Table \ref{table1} \cite{chad}. 
In terms of the time-ordered diagrams $\Sigma_\delta^a$ and $\Sigma_\delta^b$ that we discussed in the last section (Section II), it means that $\Sigma_\delta^a$ and $\Sigma_\delta^b$ are not individually invariant under the longitudinal boost $K^3$ unless $\delta =\frac{\pi}{4}$. In terms of the vacuum property, it also means that the vacuum in LFD is 
very different from the vacuum in IFD because the vacuum must be invariant under different numbers of kinematic transformations.
As summarized in Table \ref{table1} \cite{chad}, the number of kinematic generators is six in general for $0 \le \delta < \frac{\pi}{4}$ but it maximizes to seven
at $\delta = \frac{\pi}{4}$. One should note that the minimum three degrees of freedom are necessary to define the 
hyper-surface of $x^{\wh +}$ in 3+1 dimension. 
\begin{table}[h]
\caption{Kinematic and dynamic generators for different angles.}
\begin{ruledtabular}
\begin{tabular}{lcc}
Angle & Kinematic & Dynamic \\
\colrule
 $\delta =0$ &  $ {\cal K}^{\wh 1}=-J^2 $, $ {\cal K}^{\wh 2}=J^1$, $J^3$, ${\mathcal P}^1$, ${\mathcal P}^2$, ${\mathcal P}^3$ &  ${\cal D}^{\wh 1}=-K^1$, ${\cal D}^{\wh 2}=-K^2$, $K^3$, ${\mathcal P}^0$  \\
 $0 < \delta < \frac{\pi}{4}$  & $ {\cal K}^{\wh 1} $, ${\cal K}^{\wh 2} $, $J^3$, ${\mathcal P}^{\wh 1}$, ${\mathcal P}^{\wh 2}$, ${\mathcal P}_{\wh -}$ & ${\cal D}^{\wh 1}$, ${\cal D}^{\wh 2}$, $K^3$, ${\mathcal P}_{\wh +}$  \\
 $\delta = \frac{\pi}{4}$ &  $ {\cal K}^{\wh 1}=-E^1 $, $ {\cal K}^{\wh 2}=-E^2$, $J^3$, $K^3$, ${\mathcal P}^1$, ${\mathcal P}^2$, ${\mathcal P}^+$  & ${\cal D}^{\wh 1}=-F^1$, ${\cal D}^{\wh 2}=-F^2$, ${\mathcal P}^-$ \\ 
\end{tabular}
\end{ruledtabular}
\label{table1}
\end{table}

What Weinberg \cite{Weinberg} showed in IMF was to take advantage of the dynamic property of $K^3$ 
(or  the frame dependence of each individual time-ordered amplitudes) in the case of $\delta = 0$
and discard the time-ordered amplitudes connected to the current arising from the vacuum in IFD,
e.g. $\Sigma_{\delta = 0}^b = 0$ in IMF for IFD. For $\delta = \frac{\pi}{4}$, i.e., in LFD, however, 
$K^3$ is kinematic and the corresponding frame dependence of each individual time-ordered amplitudes
cannot be applied. Instead, what we take advantage of in this work is that the individual time-ordered interpolating scattering amplitudes are dependent on $\delta$ and the time-ordered amplitudes connected to the current arising from the vacuum vanishes in the limit $\delta=\frac{\pi}{4}$, e.g. $\Sigma_{\delta = \frac{\pi}{4}}^b = 0$ 
(and thus $\Sigma_{\delta = \frac{\pi}{4}}^a = \frac{1}{s-m^2}$) as we showed in Section II.
Since $K^3$ is kinematic in LFD, each individual time-ordered amplitudes are invariant under 
the longitudinal boost (or independent of the corresponding change of reference frames), e.g. 
$\Sigma_{\delta = \frac{\pi}{4}}^a = \frac{1}{s-m^2}$ or  $\Sigma_{\delta = \frac{\pi}{4}}^b = 0$ is independent of the total momentum $P_z$ of the system. In the case of 
$\Sigma_{\delta = \frac{\pi}{4}}^a = \frac{1}{s-m^2}$ or  $\Sigma_{\delta = \frac{\pi}{4}}^b = 0$, 
one may note that the individual time-ordered amplitudes are indeed invariant under all Poincar\'e transformations because the first time-ordered amplitude  takes up the whole result of the Feynman amplitude. 
In the more general case of LFD where a given physical process involves more than one non-vanishing time-ordered 
amplitudes, the individual time-ordered amplitudes are not invariant under the dynamic transformations ${\cal D}^1, \, {\cal D}^2$ and ${\mathcal P}^-$ but invariant under the kinematic transformations shown in Table \ref{table1} 
including $K^3$ in LFD. For the interpolating scattering amplitudes of $0 \le \delta < \frac{\pi}{4}$, the individual time-ordered amplitudes are not invariant under the four (instead of three) dynamic transformations but invariant under the 
six (instead of seven) kinematic transformations shown in Table \ref{table1}.  

To discuss more details of the invariance of the individual time-ordered amplitudes under the kinematic transformations,
we now revisit the previous analysis\cite{chad} on the transformations of the particle momentum components under the kinematic transformations, ${\mathcal K}^{\wh j} (j=1,2)$, and extend the analysis to include the effect of
$K^3$ transformation in order to cover the case of time-ordered amplitudes in LFD.  The transformations of the particle momentum components under other kinematic transformations such as $J^3, {\mathcal P}^{\wh j} (j=1,2)$ 
and ${\mathcal P}_{\wh -}$ are rather trivial, in the sense that the particle momentum components $P^{\wh +}$ and  $P^{\wh -}$ as well as the magnitude $|{\vec P}_\perp|$ are invariant under these transformations, and we do not discuss
them here. 

To analyze the particle momentum components under the ${\mathcal K}^{\wh j} \:(j=1,2)$ and $K^3$ transformations,
we consider both the longitudinal transformation $T_3 = e^{-i\beta_3 K^3}$ and the transverse transformation
$T_{12} = e^{-i(\beta_1 {\cal K}^{\wh 1}+\beta_2 {\cal K}^{\wh 2})}$. In particular, we follow the procedure set by Jacob and 
Wick\cite{Jacob-Wick} in defining the helicities in IFD, namely $T_3$ first and $T_{12}$ later, as the spin 
in the rest frame is initially aligned in the z-direction and the boost in the z-direction first would not change the spin direction for the procedure of defining helicities.  Although we do not involve any spin degrees of freedom in this work,
we adopt the same procedure to be consistent when we extend this work later for the spinor case.   
As discussed in Ref.\cite{carl}, this procedure of applying $T_3$ first and $T_{12}$ later is common also
in defining the LF helicities. 

Having this is mind, we first apply $T_3 = e^{-i\beta_3 K^3}$ to each of the momentum operator components $(\wh{\mu} = \wh+,\,\wh- ,\, \wh1,\, \wh2)$:
\bea
T_3^{\dagger} {\mathcal P}_{\wh{\mu}}T_3 & = & e^{i\beta_3 K^3} {\mathcal P}_{\wh{\mu}} e^{-i\beta_3 K^3} \nonumber \\
                                 & = & {\mathcal P}_{\wh{\mu}} + i\left[\beta_3 K^3, {\mathcal P}_{\wh{\mu}} \right] + \frac{i^2}{2!}\left[\beta_3 K^3, \left[\beta_3 K^3, {\mathcal P}_{\wh{\mu}} \right]\right] + \cdots
\eea
This yields
\bea
\label{T3}
T_3^{\dagger} {\mathcal P}_{\wh+}T_3 & = & \left(\cosh\beta_3 - \mathbb{S}\sinh\beta_3 \right){\mathcal P}_{\wh+}+\mathbb{C}\sinh\beta_3{\mathcal P}_{\wh-} \nonumber\\ 
T_3^{\dagger} {\mathcal P}_{\wh-}T_3 & = & \left(\cosh\beta_3 + \mathbb{S}\sinh\beta_3 \right){\mathcal P}_{\wh-}+\mathbb{C}\sinh\beta_3{\mathcal P}_{\wh+} \nonumber\\
T_3^{\dagger} {\mathcal P}^{\wh{j}}T_3 & = & {\mathcal P}^{\wh{j}}, \:\: (\wh{j}=\wh1,\,\wh2) \, .
\eea
If we apply $T_3$ to the particle momentum state $|P>$, then the particle momentum state is changed to
the state $|P'>$, where $|P>$ and $|P'>$ are the eigenstates of the operator 
${\mathcal P}_{\wh{\mu}}$ with the eigenvalues of  $P_{\wh{\mu}}$ 
and $P'_{\wh{\mu}}$, respectively. From this, one can find that the operation of 
$T_3^{\dagger} {\mathcal P}_{\wh{\mu}}T_3$ and ${\mathcal P}_{\wh{\mu}}$ to the state $|P>$ yields the eigenvalues 
$P'_{\wh{\mu}}$ and $P_{\wh{\mu}}$, respectively.
Thus, the results given in Eq.(\ref{T3}) can be translated into
\bea
\label{T3-eigen}
P'_{\wh+} & = & \left(\cosh\beta_3 - \mathbb{S}\sinh\beta_3 \right){P}_{\wh+}+\mathbb{C}\sinh\beta_3{P}_{\wh-} 
\nonumber\\ 
P'_{\wh-} & = & \left(\cosh\beta_3 + \mathbb{S}\sinh\beta_3 \right){P}_{\wh-}+\mathbb{C}\sinh\beta_3{P}_{\wh+} 
\nonumber\\
P'^{\wh{j}} & = & {P}^{\wh{j}}, \:\: (\wh{j}=\wh1,\,\wh2) \, .
\eea
This result satisfies the energy-momentum dispersion relation as it should:
\bea
P'_{\wh \mu}g^{{\wh\mu}{\wh\nu}}P'_{\wh\nu} & = & \mathbb{C}{P'}_{\wh+}^2 + 2\mathbb{S}{P'}_{\wh+} {P'}_{\wh-}-\mathbb{C}{{P'}_{\wh-}}^2-{\vec{\bf P'}_{\wh{\perp}}}^2 \nonumber \\
    & = & \mathbb{C}P_{\wh+}^2 + 2\mathbb{S}P_{\wh+}P_{\wh-}-\mathbb{C}P_{\wh-}^2-\vec{\bf P}_{\wh{\perp}}^2 
    \nonumber \\
    &= & M^2 \, .
\eea
Taking the limit $\delta \rightarrow 0$ in Eq.(\ref{T3-eigen}), we get
\bea
\label{T3IFD}
P'^0 & = & \cosh\beta_3P^0+\sinh\beta_3P^3 \nonumber\\ 
P'^3 & = & \cosh\beta_3P^3+\sinh\beta_3P^0 \nonumber\\
P'^j & = & P^j, \:\: (j=1,\, 2) \, ,
\eea
which are the usual Lorentz transformations along the $z$-direction in IFD. 
Taking the limit $\delta \rightarrow \frac{\pi}{4}$, on the other hand, we get
\bea
\label{T3LFD}
P'^- & = & {\rm e}^{-\beta_3}P^{-} \nonumber\\ 
P'^+& = & {\rm e}^{\beta_3}P^{+} \nonumber\\
P'^j & = & P^j, \:\: (j=1,\,2) \, ,
\eea
which are the expected results in LFD since $P^+$ and $P^-$ are decoupled with the corresponding scaling factors. 
This result confirms that $T_3$ is kinematical in LFD. 

After the $T_3$ (longitudinal) transformation, we now take the $T_{12}$ (transverse) transformation following the Jacob and Wick's procedure as mentioned above.  In Ref.\cite{chad}, the effect of $T_{12}$ transformation on the momentum operator ${\mathcal P}_{\wh \mu}$ was obtained 
as follows:    
\bea
\label{T12}
T_{12}^{\dagger} {\mathcal P}_{\wh+}T_{12} & = & {\mathcal P}_{\wh+} + \mathbb{S} \beta_\perp^2 \frac{(1-\cos\alpha)}{\alpha^2}{\mathcal P}_{\wh-} - \mathbb{S}\frac{\sin\alpha}{\alpha}\left(\beta_{1} {\mathcal P}^{\wh1}+\beta_{2} {\mathcal P}^{\wh2}\right) \nonumber\\ 
T_{12}^{\dagger} {\mathcal P}_{\wh-}T_{12} & = & {\mathcal P}_{\wh-}\cos\alpha + \mathbb{C} \frac{\sin\alpha}{\alpha}\left(\beta_{1} {\mathcal P}^{\wh1}+\beta_{2} {\mathcal P}^{\wh2}\right) \nonumber\\ 
T_{12}^{\dagger} {\mathcal P}^{\wh{j}}T_{12} & = & {\mathcal P}^{\wh{j}}-\beta_{j} \frac{\sin\alpha}{\alpha}{\mathcal P}_{\wh-}+\mathbb{C}\beta_{j}\frac{(\cos\alpha-1)}{\alpha^2}\left(\beta_{1} {\mathcal P}^{\wh1}+\beta_{2} {\mathcal P}^{\wh2}\right), \:\: (j=1,\,2)
\eea
where we have defined $\alpha = \sqrt{\mathbb{C}(\beta_1^2+\beta_2^2)}=\sqrt{\mathbb{C}\vec{\beta}_{\perp}^2}$.
It is interesting to note that this result indicates a dramatic difference in the outcome of the particle momentum
after the application of the kinematic transformation $T_{12} = e^{-i(\beta_1 {\cal K}^{\wh 1}+\beta_2 {\cal K}^{\wh 2})}$ to the particle in the rest frame between IFD ($\delta=0$) and LFD ($\delta=\pi/4$).
The particle of mass $M$ in the rest frame (i.e. $P^0=M, {\vec P}=0$) has the interpolating momentum components 
given by $P_{\wh+} = M\cos\delta, P_{\wh-}=M\sin\delta, \vec{\bf P}_{\wh{\perp}}=0$. 
If we write the interpolating momentum components with the prime notation after the $T_{12}$ transformation, we get 
\bea
\label{T12a}
P'_{\wh+} & = & M\left[\cos\delta + \mathbb{S}\vec{\beta}_\perp^2\frac{\left(1-\cos\alpha\right)}{\alpha^2} \sin\delta\right] \nonumber \\
P'_{\wh-} & = & M\sin\delta\cos\alpha \nonumber \\
P'^{\wh{j}}  & = & -M\beta_j\sin\delta\frac{\sin\alpha}{\alpha} \qquad (\wh{j}=\wh{1},\,\wh{2}) \, ,
\eea
which shows that the particle can gain some longitudinal momentum although the transformation $T_{12}$ is transversal
and the amount of the gained longitudinal momentum depends on the interpolating angle $\delta$.  
In IFD ($\delta = 0$), the particle in the rest frame remains in the rest frame since $T_{12}$ is just a transverse rotation:
i.e. $P'^0 = M, {\vec P'}=0$. However, in LFD ($\delta= \frac{\pi}{4}$), the result given by Eq.(\ref{T12}) can be written as
\bea
P'^{-} &=&  \frac{M}{\sqrt{2}}\left(1+\frac{\vec{\beta}^2_{\perp}}{2}\right) \nonumber \\
P'^{+} &=&  \frac{M}{\sqrt{2}} \nonumber \\
P'^{j} & = & -\frac{M}{\sqrt{2}}\beta_{j} \, , \, (j=1,2).
\eea
From this, we find the energy and longitudinal momentum components are related to the transverse momentum
$\vec{\bf P'}_\perp= - M \vec{\bf {\beta}}_{\perp}/ \sqrt{2}$, i.e.
\bea
\label{rel-LF}
P'^0 & = & M+\frac {\vec{\bf P'}_\perp^2}{2M} \nonumber \\
P'^3 & = & -\frac{\vec{\bf P'}_\perp^2}{2M} \, 
\eea
which shows that the particle gains the longitudinal momentum $ -\frac{\vec{\bf P'}_\perp^2}{2M}$ while the particle
is transversely boosted by $T_{12}=  e^{i(\beta_1 E^1+\beta_2 E^2)}$. One should note that the LF transverse boosts
$E^1 = (J^2 + K^1)/\sqrt{2}$ and $E^2 = (K^2 - J^1)/\sqrt{2}$ involve not only $K^1, K^2$ (ordinary transverse boosts)
but also $J^1, J^2$ (ordinary transverse rotation) so that the particle's moving direction cannot be kept just in the transverse direction while the particle is transversely boosted. 
This yields the momentum in the longitudinal direction as well as in the transverse direction. 
 It is also interesting to note that the relativistic energy-momentum dispersion relation works although the particle energy takes a non-relativistic form:  
\be
\left(P^0\right)^2-\vec{\bf P}^2 = \left(M+\frac{\vec{\bf P}_\perp^2}{2M}\right)^2-\vec{\bf P}_\perp^2-\left(-\frac{\vec{\bf P}_\perp^2}{2M}\right)^2 = M^2 \, .
\ee
This may be regarded as another distinguishing feature of the LFD.  
 
We now apply the $T_{12}$ transformation subsequently after we do the $T_3$ transformation in order to combine the longitudinal boost and the transverse kinematic transformations, i.e. $T_K = T_3T_{12} = e^{-i\beta_3K^3}e^{-i\left(\beta_1{\cal K}^{\wh 1}+\beta_2{\cal K}^{\wh 2} \right)}$.
This allows not only the transformation of the unprimed $P_{\wh{\mu}}$ to primed $P'_{\wh{\mu}}$ but also
the subsequent transformation from the primed four-momentum $P'_{\wh{\mu}}$ to the double-primed four-momentum $P''_{\wh{\mu}}$ of  the particle that we consider. Under the $T_K$ transformation, we get
\bea
{\mathcal P}''_{\wh{\mu}} & = & T_K^{\dagger} {\mathcal P}_{\wh{\mu}}T_K \nonumber \\
               & = & T_{12}^{\dagger}\left(T_3^{\dagger} {\mathcal P}_{\wh{\mu}} T_3 \right) T_{12} \nonumber \\
               & = & T_{12}^{\dagger} {\mathcal P}'_{\wh{\mu}} T_{12} \, .
          \eea
From this, we get the following general transformation relations:
\bea
\label{general_kinematic}
P''_{\wh+} & = & \left(\cosh\beta_3 - \mathbb{S} \cos\alpha \sinh\beta_3\right )P_{\wh+} \nonumber \\
                         & + & \left[\left(1-\mathbb{S}^2\cos\alpha \right)\sinh\beta_3 + \mathbb{S}\left(1-\cos\alpha \right)\cosh\beta_3 \right]\frac{\vec{\beta}^2_\perp}{\alpha^2}P_{\wh-} \nonumber \\
                         & - & \mathbb{S} \frac{\sin\alpha}{\alpha}\left(\beta_1 P^{\wh1}+\beta_2 P^{\wh2} \right)  
                         \nonumber\\
P''_{\wh-} & = & \mathbb{C} \cos\alpha\sinh\beta_3 P_{\wh+} + \cos\alpha\left(\cosh\beta_3 + \mathbb{S}\sinh\beta_3\right )
P_{\wh-} \nonumber \\
                         & + & \mathbb{C} \frac{\sin\alpha}{\alpha}\left(\beta_1 P^{\wh1}+\beta_2 P^{\wh2} \right)  \nonumber \\
P''^{\wh{j}} & = & P^{\wh{j}} - \mathbb{C}\beta_j\frac{\sin\alpha}{\alpha}\sinh\beta_3 P_{\wh+} - \beta_j \frac{\sin\alpha}{\alpha}\left(\cosh\beta_3 + \mathbb{S}\sinh\beta_3\right ) P_{\wh-} \nonumber \\
                           & + & \mathbb{C}\beta_j \frac{(\cos\alpha-1)}{\alpha^2}\left(\beta_1 P^{\wh{1}}+\beta_2 P^{\wh{2}} \right) \, ,
\eea
which of course satisfy the dispersion relation as expected:
\bea
M^2 & = & 
\mathbb{C} {P''_{\wh+}}^2 + 2\mathbb{S}P''_{\wh+}P''_{\wh-} - {\mathbb{C}P''_{\wh-}}^2 -{\vec{\bf P''}_{\wh{\perp}}}^2
\nonumber \\
 & = & 
 \mathbb{C}P_{\wh+}^2 + 2\mathbb{S}P_{\wh+}P_{\wh-}-\mathbb{C}P_{\wh-}^2-\vec{\bf P}_{\wh{\perp}}^2 \, .
\eea

In the IFD limit, $\delta \rightarrow 0$, we note that
$\alpha^2 \rightarrow (\beta_{1}^2+\beta_{2}^2) = \vec{\bf \beta}_{\perp}^2$ and 
get
\bea
\label{IFD}
P''^0 & = & \cosh\beta_3 P^0 + \sinh\beta_3 P^3 \nonumber \\
P''^3 & = & \cos\beta_{{\perp}} \sinh\beta_3 P^0 + \cos\beta_{{\perp}} \cosh\beta_3 P^3 +\frac{\sin\beta_{{\perp}}}{\beta_{{\perp}}}\left(\beta_{1} P^{1}+\beta_{2} P^{2} \right) \nonumber \\
P''^{{j}} & = & P^{{j}} - \beta_{{j}}\frac{\sin\beta_{{\perp}}}{\beta_{{\perp}}}\left(\sinh\beta_3 P^0 + \cosh\beta_3 P^3\right) \nonumber \\
                           & + & \beta_{{j}} \frac{\left(\cos\beta_{{\perp}}-1\right)}{\beta_{{\perp}}^2}\left(\beta_{1} P^{1}+\beta_{2} P^{2} \right)  \, ,
                      \eea
where $\beta_{\perp}=\sqrt{{\vec{\bf \beta}_{\perp}^2}}$.
Here, the transverse vector $\vec{\bf \beta}_{\perp}=(\beta_{1},\, \beta_{2})$ can be represented by $\vec{\bf \beta}_{\perp}=\theta (\hat{\bf z}\times\hat{\bf n}_\perp)$ 
defining the angle $\theta$ and the rotation axis as the unit transverse vector $\hat {\bf n}_\perp=(n_{1},\,n_{2})$ 
because the kinematic transformations ${\mathcal K}^{\wh 1}$ and ${\mathcal K}^{\wh 2}$ are nothing
but the ordinary transverse rotations $-J^2$ and $J^1$, respectively, in IFD.
Since  $\hat {\bf z}\times \hat {\bf n}_\perp = -n_{2}\hat {\bf x}+n_{1} \hat{\bf y} = (-n_{2},\, n_{1})$,  
one may identify $\beta_{1} = -\theta n_{2}$ and $\beta_{2} = \theta n_{1}$ to rewrite Eq.(\ref{IFD}) as 
\bea
\label{IFD1}
P''^0  &=&  \cosh\beta_3 P^0 + \sinh\beta_3 P^ 3  \nonumber \\
P''^ 3 &=&  \cos\theta\left(\sinh\beta_3 P^0 + \cosh\beta_3 P^3\right)
+\sin\theta\left(\hat{\bf z}\times\hat{\bf n}_\perp \right)\cdot\vec{\bf P}_\perp 
\nonumber \\
\vec{\bf P''}_\perp & = & \vec{\bf P}_\perp 
- (\hat{\bf z}\times\hat{\bf n}_\perp)\sin\theta\left(\sinh\beta_3 P^0 + \cosh\beta_3 P^3\right) \nonumber \\
&+& (\hat{\bf z}\times\hat{\bf n}_\perp)(\cos\theta-1)(\hat{\bf z}\times\hat{\bf n}_\perp)\cdot \vec{\bf P}_\perp \, .
\eea
Taking $\hat {\bf n}_\perp = \hat {\bf y}$ (i.e. $\hat{\bf z}\times\hat{\bf n}_\perp = - \hat {\bf x}$), we have
\bea
\label{IFD2}
P''^0 & = & P'^0 = \cosh\beta_3 P^0 + \sinh\beta_3 P^{3} \nonumber \\
P''^1 & = & -\sin\theta P'^3 + \cos\theta P'^1 = -\sin\theta\left(\sinh\beta_3 P^0 + \cosh\beta_3 P^3\right) + \cos\theta P^1
\nonumber \\
P''^2 & = & P'^2 = P^2 \nonumber \\ 
P''^3 & = & \cos\theta P'^3 + \sin\theta P'^1 = \cos\theta\left(\sinh\beta_3 P^0 + \cosh\beta_3 P^3\right) +\sin\theta P^1
\, ,
                      \eea
where the boost in $\hat{\bf z}$ direction and the subsequent rotation around $\hat{\bf y}$ axis are manifest. 

\begin{figure}[ht]
\begin{center}
\includegraphics[scale=.8]{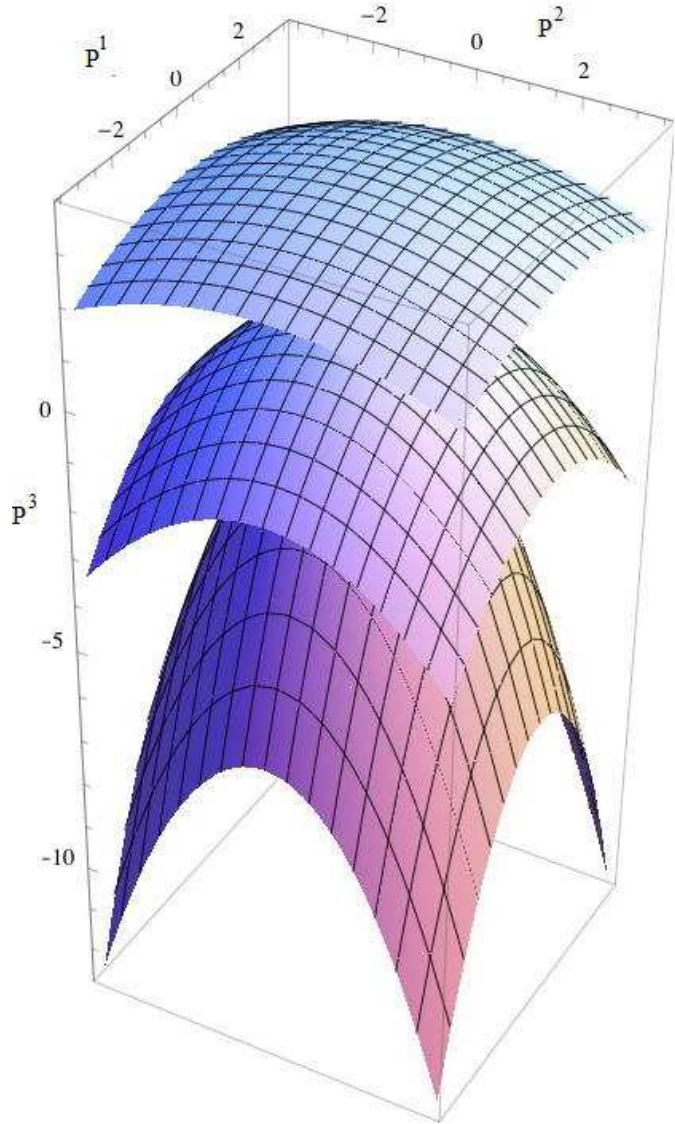}
\caption{General kinematic transformation on a fixed interpolating front.}
\label{fig:3} 
\end{center}      
\end{figure}

Next, we consider the other extreme that corresponds to the LFD, $\delta = \frac{\pi}{4}$.  As $\delta \rightarrow \frac{\pi}{4}$,$\,\, \alpha \rightarrow 0$ and it leads to the following limits for the expressions 
that appear in the different components of momentum given by Eq.(\ref{general_kinematic}):
\bea
\left(1-\mathbb{S}^2\cos\alpha \right)\frac{\vec\beta_\perp^2}{\alpha^2} & \rightarrow & \frac{\vec{\bf \beta}_{\perp}^2}{2} \nonumber \\
\frac{\left(1-\cos\alpha \right)}{\alpha^2} & \rightarrow & \frac{1}{2} \nonumber \\
\frac{\sin\alpha}{\alpha} & \rightarrow & 1 \, .
\eea
Using the usual LFD notations, we thus get
\bea
P''^- & = & e^{-\beta_3}P^-+\frac{e^{\beta_3}\vec{\bf \beta}_\perp^2}{2}P^+-\vec{\bf \beta}_\perp \cdot \vec{\bf P}_\perp \nonumber \\
P''^+ & = & e^{\beta_3} P^+ \nonumber \\
\vec{\bf P''}_\perp & = & \vec{\bf P}_\perp - e^{\beta_3} \vec{\bf \beta}_\perp P^+ \, ,
\eea
which satisfies the LF dispersion relation as expected
\be
2P''^+P''^- -\vec{\bf P''}_\perp^2 = 2P^+P^- -\vec{\bf P}_\perp^2 = M^2 \, .
\ee
In the case that the particle is at rest in the unprimed frame, i.e. 
\bea
P^+ & = & P^- = \frac{M}{\sqrt {2}} \nonumber \\
\vec {\bf P}_\perp & = & 0 \, ,
\eea
we obtain 
\bea
P''^{-} & = & \frac{M}{\sqrt{2}}\left(e^{-\beta_3}+e^{\beta_3}\frac{\vec {\beta}_\perp^2}{2} \right) \nonumber \\
P''^{+} & = & \frac{M}{\sqrt{2}}e^{\beta_3} \nonumber \\
\vec{\bf P}''_\perp & = & - \frac{M}{\sqrt{2}}\vec{\beta}_\perp e^{\beta_3} \, ,
\eea 
which can be translated into
\bea
P''^{0} & = & M \cosh\beta_ 3 +\frac{M}{4}\vec {\beta}_\perp^2 e^{\beta_3} \nonumber \\
P''^{3} & = & M \sinh\beta_3 - \frac{M}{4}\vec {\beta}_\perp^2 e^{\beta_3} \nonumber \\
\vec{\bf P''}_\perp & = & - \frac{M}{\sqrt{2}}\vec{\beta}_\perp e^{\beta_3} \, .
\eea 
From this, we may extend the relation between the energy and the transverse momentum 
(as well as between the longitudinal momentum and the transverse momentum) given by Eq.(\ref{rel-LF})  as 
\bea
\label{extend-rel-LF}
P''^{0} & = & M \cosh\beta_ 3 +\frac{\vec{\bf P''}_\perp^2}{2M} e^{-\beta_3} \nonumber \\
P''^{3} & = & M \sinh\beta_3 - \frac{\vec{\bf P''}_\perp^2}{2M} e^{-\beta_3} \, .
\eea 
For  $\beta_3 = 0$, this equation is reduced to Eq.(\ref{rel-LF}). As we explained about  Eq.(\ref{rel-LF}), the gained longitudinal momentum is correlated with the transverse momentum due to the kinematic transformation $T_{12}=  e^{i(\beta_1 E^1+\beta_2 E^2)}$
in such a way that a paraboloid shape of surface (note $P''^3= -\frac{\vec{\bf P''}_\perp^2}{2M}$ for $\beta_3 = 0$) can be drawn for the gained momentum components  in the momentum space as shown in Ref.\cite{chad}. 
In the case $\beta_3 \neq 0$, we find that the similar shapes of paraboloids can be drawn. However, the corresponding
paraboloids are shifted in the longitudinal direction as $\beta_3$ gets more positive values and the curvatures of the corresponding paraboloids get modified as shown in Fig.\ref{fig:3}. This plot shows three surfaces corresponding to three different values of $\beta_3 = 0, 1, 2$, with 
the momenta scaled by the mass of the particle, i.e. ${\vec p} = {\vec P''}/M$, in the range $-4 < \vec{p}_\perp < 4$ and $-12 < p_z < 4 $. For the positive values of $\beta_3$ as shown in Fig.\ref{fig:3}, the paraboloid of $\beta_3 = 0$ is shifted
to upwards in $p_z$ and gets flattened due to the factors given by ${\rm sinh} \beta_3$ and $e^{-\beta_3}$ in Eq.(\ref{extend-rel-LF}), respectively. The top point of each paraboloid corresponds to the momentum gained by the $T_3 = 
e^{-i \beta_3 K^3}$ transformation in IFD (see Eq.(\ref{T3IFD})). Although the particle at rest stays at rest in IFD
when only the kinematic transformation $T_{12}$ (i.e. the ordinary transverse rotation in IFD) is applied, the longitudinal boost $T_3$ is dynamical in IFD so that it can generate the longitudinal momentum of the particle.
However, in LFD, both $T_{12}$ and $T_3$ are kinematic transformations and the entire momentum region of $\vec p$
can be covered by these kinematic transformations.

\section{Application of Transformations on Interpolating Scattering Amplitudes}
\label{sec.04}

In the previous sections, we discussed that the scattering amplitude in Fig.\ref{fig:1} has two non-vanishing time-ordered contributions in an arbitrary interpolating angle for the range $0 \le \delta < \frac{\pi}{4}$ including IFD ($\delta=0$)
while in LFD ($\delta = \frac{\pi}{4}$) only the contribution of the first diagram Fig.\ref{fig:2}a survives.
We now apply the transformations of the particle momenta that we obtained in the last section, Section~\ref{sec.03}, to the scattering amplitudes and discuss a quantitative measure on the invariance of the individual time-ordered amplitudes under the kinematic transformations. 

In order to see this in an arbitrary  interpolating angle, let us first consider the expression for $D_{+}$ found in Eq. (\ref{D+D-}) under the transverse kinematic boost $T_{12}$, i.e.
\be
D'_+  =  P'_{\wh+} + \frac{\mathbb{S} q'_{\wh-} - \omega_q'}{\mathbb{C}},
\ee
where the prime indicates the transformed frame variables via ${\mathcal P}'_{\wh+} = T_{12}^{\dagger}{\mathcal P}_{\wh+}T_{12}$, etc.
This quantity $D'_{+}$ expresses the difference between the interpolating angle energies of $P'_{\wh +}$ and $q'_{\wh +}$
for the first diagram Fig.\ref{fig:2}a. Under $T_{12}$ (see Eq. (\ref{T12})), we get 
\bea
D'_+  & = &  P_{\wh+} + \mathbb{S}\frac{(\beta_{\wh1}^2+\beta_{\wh2}^2)}{\alpha^2}(1-\cos\alpha)P_{\wh-}-\mathbb{S}\frac{\sin\alpha}{\alpha}(\beta_{\wh1}P^{\wh1}+\beta_{\wh2}P^{\wh2})\nonumber \\
      & - &  \frac{\mathbb{S}}{\mathbb{C}}\left[q_{\wh-}\cos\alpha+\mathbb{C}\frac{\sin\alpha}{\alpha}(\beta_{\wh1}P^{\wh1}+\beta_{\wh2}P^{\wh2})\right] - \frac{\omega_q'}{\mathbb{C}} \nonumber \\
      & = &  P_{\wh+} + \frac{\mathbb{S}}{\mathbb{C}}q_{\wh-} - \frac{\omega_q'}{\mathbb{C}},
\eea 
where we used  $\alpha = \sqrt{\mathbb{C}(\beta_{\wh1}^{2}+\beta_{\wh2}^{2})} $ and the momentum conservation $P_{\wh-} = q_{\wh-}$.
This means that if $\omega'_q = \omega_q$ as defined by Eq. (\ref{omegaq}), then $D'_{+} = D_{+}$ and the first term by itself is invariant under $T_{12}$. We may use the solution in terms of $q_{\wh+}$ of the quadratric equation for the dispersion relation and show $\omega'_q = \omega_q$: i.e.
\be
q_{\wh+} = \frac{\omega_q-\mathbb{S}q_{\wh-}}{\mathbb{C}} \Rightarrow \omega_q = \mathbb{C}q_{\wh+}+\mathbb{S}q_{\wh-}
\ee 
so that 
\bea
\omega'_{q} & = & \mathbb{C}q'_{\wh+}+\mathbb{S}q'_{\wh-}\nonumber \\
            & = & \mathbb{C}q_{\wh+}+\mathbb{S}q_{\wh-} = \omega_q \, ,
            \eea
according to Eq. (\ref{T12}). It is now manifest that $D_{+}$ by itself is invariant under $T_{12}$. Similar manifestation can be obtained for $D_{-}$ for the second diagram Fig.\ref{fig:2}b. 

Now, we apply the longitudinal boost $T_3$ to the interpolating time-ordered amplitudes. 
As we have already discussed in Section \ref{sec.03}, the longitudinal boost $K_3$ is dynamical for any $\delta$ in the range $0\le \delta < \frac{\pi}{4}$ and becomes kinematical only at $\delta=\frac{\pi}{4}$. 
To exhibit this feature quantitatively, we show Fig. 4 which plots $\Sigma^a_\delta$ and  $\Sigma^b_\delta$ as functions of the initial particle total momentum $({\vec p}_{\wh 1}+{\vec p}_{\wh 2})\cdot {\hat z} = P_z$ while $({\vec p}_{\wh 1}+{\vec p}_{\wh 2})\cdot {\hat x} = 0$ and $({\vec p}_{\wh 1}+{\vec p}_{\wh 2})\cdot {\hat y} = 0$ for convenience, 
as well as the interpolation angle $\delta$.  The ranges of $\delta$ and $P_z$ are taken as  $0 \le \delta < \frac{\pi}{4}$ and $-4 \le P_z \le 4$ in some unit of energy, e.g. GeV, respectively. For illustrative purpose, we took $s=2$ and $m=1$ using the same energy unit.  
\begin{figure}[h]
\begin{center}
\includegraphics[scale=.7]{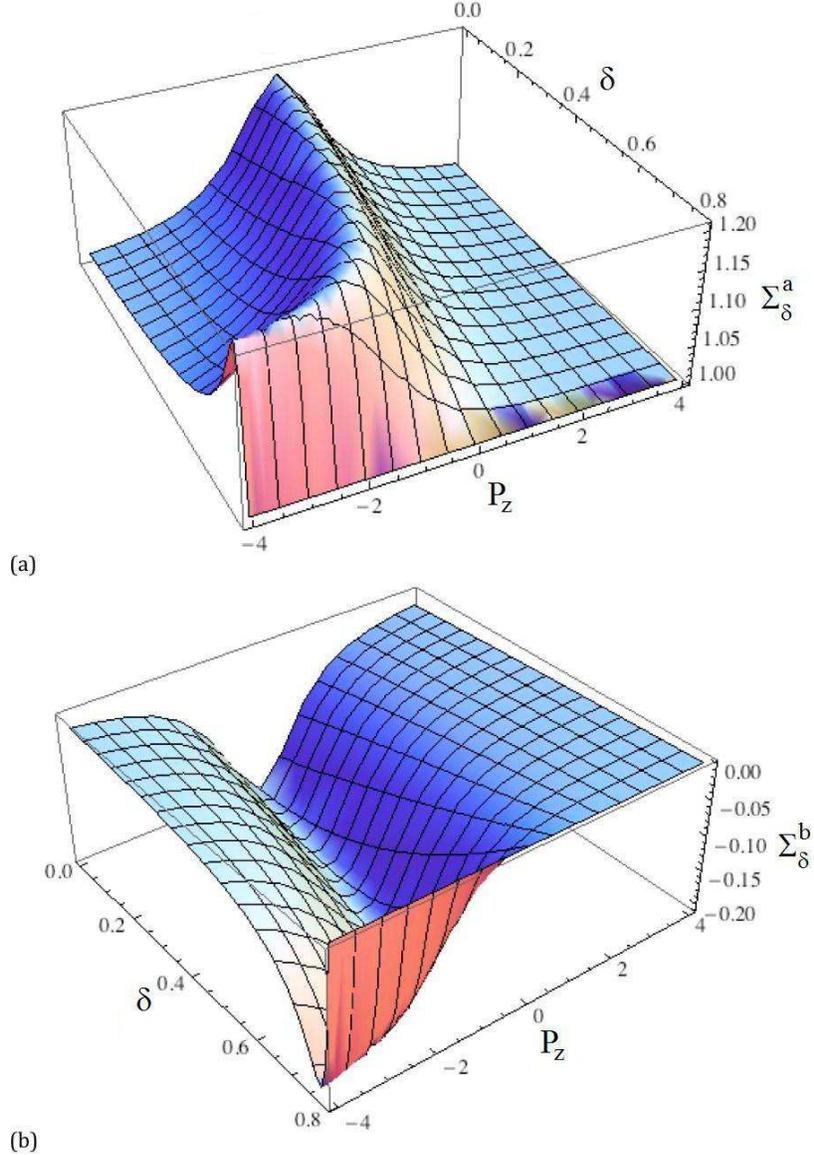}
\caption{Interpolating Amplitudes}
\label{fig:4} 
\end{center}      
\end{figure}
As clearly shown in Fig. 4, the contributions from $\Sigma^a_\delta$ and  $\Sigma^b_\delta$ are such that the sum of them yields a constant, independent of $P_z$ and $\delta$. For $\delta=0$, $\Sigma^a_\delta$ and $\Sigma^b_\delta$ has the maximum and the minimum, respectively, at $P_z=0$. For $\delta=\frac{\pi}{4}$, $\Sigma^a_\delta$ is the whole answer and $\Sigma^b_\delta=0$. For positive values of momentum, $P_z>0$, the amplitudes $\Sigma^a_\delta$ and $\Sigma^b_\delta$ show a smooth behaviour (see also Appendix), while for negative values of $P_z$ we observe the presence of a $J$-shaped curve in the peak of $\Sigma^a_\delta$ matched by a similar $J$-shaped curve in the valley of $\Sigma^b_\delta$. We find that this $J$-shaped curve of maximum/minimum is given by the function 
$P_z=-\sqrt{\frac{s(1-\mathbb{C})}{2\mathbb{C}}}$. This $J$-shaped curve is plotted in Fig. \ref{fig:5}. On this $J$-shaped curve, a stable maximum and minimum of $\Sigma_\delta^a$ and $\Sigma_\delta^b$, respectively, is present for the negative values of momentum $P_z$: i.e. 
\begin{eqnarray}
\label{J-shape-Amps}
\Sigma_\delta^a &=&\frac{1}{2m(\sqrt{s}-m)} \, , \nonumber\\
\Sigma_\delta^b &=&-\frac{1}{2m(\sqrt{s}+m)} \, , \nonumber\\
\Sigma_\delta^a &+& \Sigma_\delta^b = \frac{1}{s-m^2} \, .
\end{eqnarray}
 
\begin{figure}[ht]
\begin{center}
\includegraphics[scale=1.0]{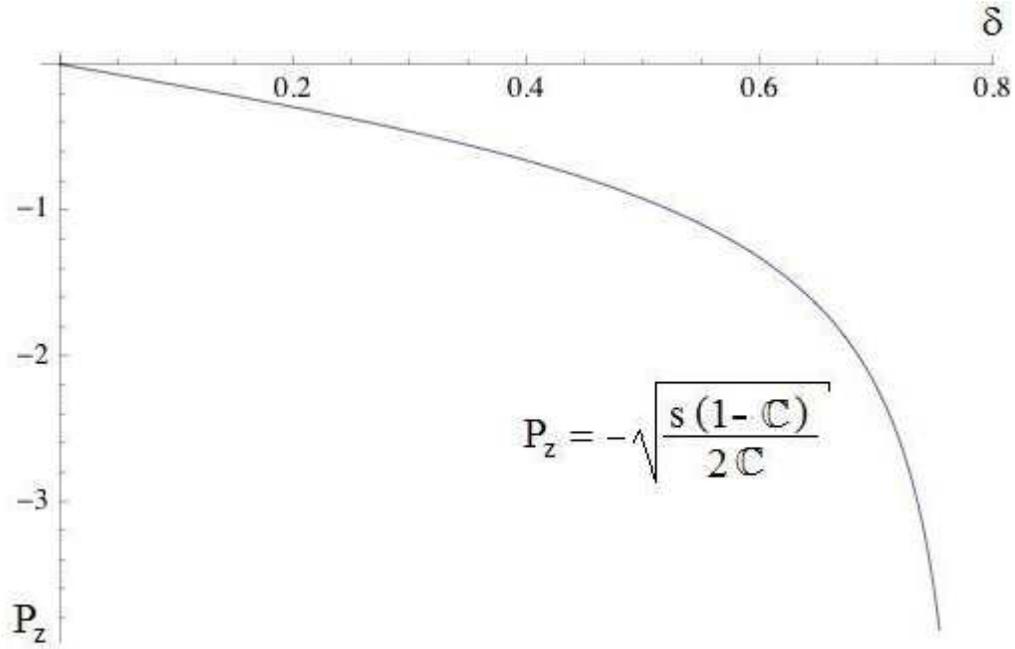}
\caption{$J$-shaped curve of maximum/minimum for $\Sigma_\delta^{a}$ and $\Sigma_\delta^{b}$}
\label{fig:5} 
\end{center}      
\end{figure}

One interesting point to observe in this $J$-shaped curve for negative values of momentum $P_z$ is that it is stable in the peak as well as in the valley as it is independent of the mass and does not vanish as the momentum goes to the negative infinity. Thus, if the limit $\delta \rightarrow \frac{\pi}{4}$ is taken in the exact correlation with $P_z$ given by the $J$-shaped curve, i.e. 
$P_z=-\sqrt{\frac{s(1-\mathbb{C})}{2\mathbb{C}}} \stackrel{\mathbb{C}\rightarrow 0} \longrightarrow -\infty$ , then the connected contribution to the current arising from the vacuum $\Sigma_{\delta\rightarrow \frac{\pi}{4}}^b$ does not vanish but remains as a nonzero constant, i.e. $-\frac{1}{2m (\sqrt{s}+m)} =-\frac{1}{2(\sqrt{2}+1)}\approx -0.207$, although this nonzero constant (i.e. the minimum of $\Sigma_{\delta\rightarrow \frac{\pi}{4}}^b$) is cancelled by the same magnitude of the constant (i.e. the maximum of $\Sigma_{\delta\rightarrow \frac{\pi}{4}}^a$) given by $\frac{1}{2m (\sqrt{s}-m)} =\frac{1}{2(\sqrt{2}-1)}\approx 1.207$ to yield the total amplitude $\frac{1}{s-m^2}=1$.

This may clarify the prevailing notion of the equivalence between IFD and LFD in the IMF since it works for the  
limit of $P_z \rightarrow \infty$ but requires a great caution in the limit of $P_z \rightarrow -\infty$.
Although the IFD in IMF is entirely symmetric between $P_z = \infty$ and $P_z = -\infty$, there is treacherous point 
$P_z = -\infty$ in LFD.  As far as the limit of $P_z = -\infty$ is taken off from the $J$-shaped curve, i.e., {\it without} the specific correlation $P_z=-\sqrt{\frac{s(1-\mathbb{C})}{2\mathbb{C}}} \stackrel{\mathbb{C}\rightarrow 0} \longrightarrow -\infty$, then our result of $\Sigma^b_{\delta = \frac{\pi}{4}}=0$ is valid. However, if the limit of $P_z = -\infty$ is taken exactly {\it with} this particular correlation, then the result $\Sigma^b_{\delta = \frac{\pi}{4}}=0$ is not correct but should be modified to be the nonzero minimum value of $\Sigma^b_{\delta= \frac{\pi}{4}} = - \frac{1}{2m (\sqrt{s}+m)} \neq 0$.
In this sense, the $J$-shaped curve which we find in this work is singular. Nevertheless, even in this case, 
the sum of the two amplitudes $\Sigma^a_{\delta = \frac{\pi}{4}}+\Sigma^b_{\delta = \frac{\pi}{4}}$ remains invariant
as it should be. 

\section{Conclusions}
\label{sec.05}

In the present work, we discussed the fundamental aspects of the time-ordered scattering amplitudes in relativistic Hamiltonian dynamics. Using the interpolating angle between IFD and LFD, we presented a simple but clear example of interpolating scattering amplitudes and demonstrated a physical meaning of kinematical transformations introduced often formally in the stability group of Poincar\'e transformations. We confirmed the well-known IMF result\cite{Weinberg} for the IFD and extended it for any arbitrary interpolating angle $0 \le \delta < \frac{\pi}{4}$. We also showed that the disappearance of the connected contributions to the current from the vacuum in LFD is independent of the reference frame and should be distinguished from the usual IMF result. We demonstrated that the longitudinal boost $K^3$ joins the stability group only in the LFD. We did this not only using explicit expressions of kinematic transformation effects on the fundamental dynamical variables of physical momenta but also discussing the interpolating time-ordered scattering amplitudes. The addition of $K^3$ in the stability group is a great advantage of LFD in hadron phenomenology\cite{carl}.

Computing the individual time-ordered amplitudes for the whole range of total momentum $P_z$ and the interpolating angle $\delta$, we showed not only the invariance of the sum of time-ordered amplitudes but also the behavior of each individual time-ordered amplitudes (see Fig. \ref{fig:4}). Our work demonstrates a rather clear distinction between the well-known IMF result in IFD and the LFD result on the disappearance of the connected contribution to the current from the vacuum. 
Our result exhibits the $J$-shaped curve given by $P_z = -\sqrt{\frac{s(1-\mathbb{C})}{2\mathbb{C}}}$ which reminds a treacherous zero-mode issue in LFD. 
The $J$-shaped curve provides a correlation between the total momentum $P_z$ and the interpolation angle $\delta$.
It traces the maximum of the time-ordered amplitude $\Sigma^a_{0 \le \delta < \frac{\pi}{4}}$ as well as the minimum of the time-ordered amplitude $\Sigma^b_{0 \le \delta < \frac{\pi}{4}}$.  Thus,  
if one takes the interpolating angle to the limit of $\frac{\pi}{4}$ in an exact correlation with the limit $P_z \rightarrow -\infty$ following the $J$-shaped curve, then one should be careful not to miss the contribution from the minimum value of $\Sigma^b_{0\le \delta < \frac{\pi}{4}}$ which must be cancelled by the maximum value of $\Sigma^a_{0\le \delta < \frac{\pi}{4}}$. Although our work is limited to a simple example without spins or any other degrees of freedom except the particle momenta, the results seem to offer interesting and significant aspects of the relativistic Hamiltonian dynamics which interpolates between IFD and LFD.   

\begin{acknowledgments}
We thank Stan Brodsky for his interest and valuable comments on this work.
This work is supported by the U.S. Department of energy (DE-FG02-03ER41260).
ATS acknowledges partial support from CNPq-Conselho Nacional de Desenvolvimento Cient\'{\i}fico e Tecnol\'ogico (Proc.  201.902/2010-9), and the hospitality of Physics Department of North Carolina State University for the sabbatical leave as well as the hospitality of Physics and Engineering Department, Southern Adventist University.  
\end{acknowledgments}

\appendix
\section{Interpolating Scattering Amplitudes in Infinite Momentum Frame}
\label{sec:IMF-extension}

As we discussed in Section \ref{sec.02}, we can rewrite the interpolating time-ordered amplitudes in the same form as in the IFD by changing the superscript $0$ (i.e. the energy) to superscript $\wh+$ as well as multiplying an overall factor $\mathbb{C}$. Then it follows that interpolating amplitudes become IFD amplitudes as $\mathbb{C} \rightarrow 1$. In the LFD case as $\mathbb{C} \rightarrow 0$, the fraction $\frac{1}{P^+-q^+} \rightarrow \infty$ due to the conservation of $P^+ = q^+$ but the multiplication of zero and infinity makes the finite $\frac{1}{s-m^2}$ just from the first diagram alone, while the second diagram vanishes since the denominator $P^+ + q^+$ is nonzero.  
The disappearance of the connected contributions to the current arising from the vacuum at $\mathbb{C} =0$ (LFD), i.e. $\Sigma^b_{\delta=\pi/4}=0$, should be distinguished from the similar disappearance of $Z$-graph in the IMF at $\mathbb{C}=1$ (IFD). In this Appendix, we apply the longitudinal boost $T_3$ (see Eq.(\ref{T3})) and take a specific limit
to an infinite momentum frame, viz. $(P_{z},\,q_{z}) \equiv (P^3,\,q^3) \rightarrow \infty$, in order to discuss more details
of the disappearance of the connected contributions for the entire range of the interpolation angle $0 \le \delta \le \frac{\pi}{4}$. 

First of all, let us consider the case of the IFD (see Eq.(\ref{IFD_amp})), where the longitudinal component of interest is $P_{\wh-} = P_z \equiv P^3$, etc. The time-ordered diagram of Fig.\ref{fig:1} is dependent on the reference frame:
\be
\label{IFD_aa}
\Sigma_{\rm IFD}^{a} = \frac{1}{2q^0}\left(\frac{1}{P^0 - q^0} \right).
\ee
From the dispersion relation $q^2 = m^2$, the expansion of $q^0$ for the IMF is given by 
\bea
\label{q}
q^0 & = & \sqrt{\vec{\bf q}^2 + m^2}  =  \sqrt{q_z^2+\vec{\bf q}_{\perp}^2+m^2}, \nonumber \\
    & = & q_z\left\{1+\frac{\vec{\bf q}_{\perp}^2+m^2}{2q_z^2}+{\mathcal O}\left(\frac{1}{q_z^4}\right)\right\} \, .
\eea
Similarly, from the dispersion relation $P^2 = s$, the expansion of $P^0$ for the IMF is given by 
\bea
\label{P}
P^0 & = & \sqrt{\vec{\bf P}^2 + s}  =  \sqrt{P_z^2+\vec{\bf P}_{\perp}^2+s}, \nonumber \\
    & = & P_z\left\{1+\frac{\vec{\bf P}_{\perp}^2+s}{2P_z^2}+{\mathcal O}\left(\frac{1}{P_z^4}\right)\right\}.
\eea
Substituting Eq. (\ref{q}) and Eq. (\ref{P}) into Eq. (\ref{IFD_aa}), we get
\be
\label{IFD_a1}
\Sigma_{\rm IFD}^{a} = \frac{1}{2q_z\left\{1+\frac{\vec{\bf q}_{\perp}^2+m^2}{2q_z^2}+{\mathcal O}\left(\frac{1}{q_z^4}\right)\right\}}\left\{\frac{1}{P_z-q_z+\frac{\vec{\bf P}_{\perp}^2+s}{2P_z}-\frac{\vec{\bf q}_{\perp}^2+m^2}{2q_z}+{\mathcal O}\left(\frac{1}{q_z^3},\,\frac{1}{P_z^3}\right)} \right\}.
\ee
Due to the three-momentum conservation, $P_z = q_z$ and $\vec{\bf P}_{\perp} = \vec{\bf q}_{\perp}$, the result (\ref{IFD_a1}) in the IMF limit yields
\be
\label{IFD_a2}
\Sigma_{\rm IFD}^{a}  =  \frac{1}{2q^0}\left(\frac{1}{P^0 - q^0} \right) 
                      \xrightarrow{P_z = q_z \rightarrow \infty}  \frac{1}{s-m^2}.
\ee
Likewise, for the diagram of Fig.\ref{fig:2}b, we get
\be
\label{IFD_b2}
\Sigma_{\rm IFD}^{b}  =  \frac{1}{2q^0}\left(\frac{1}{P^0 + q^0} \right) 
                      \xrightarrow{P_z = q_z \rightarrow \infty} 0.
\ee
This reveals that the results (\ref{IFD_a2}) and (\ref{IFD_b2}) are frame-dependent. 

Next, we consider what happens in the LFD case, where we have $P_{-} = P^{+}$ and $q_{-} = q^{+}$. 
Independent of reference frames, i.e. regardless of the $P_z$ value, the result is given by
\bea
\label{LFD1ab}
\Sigma_{\rm LFD}  & \equiv & \Sigma_{\rm LFD}^a =  \frac{1}{2q^+}\left(\frac{1}{P^- - \frac{\vec{\bf q}_{\perp}^2+m^2}{2q^+}} \right)\nonumber\\
                  & = & \frac{1}{2q^+P^--(\vec{\bf q}_{\perp}^2+m^2)}.
\eea 
Since $q^+ = P^+, \, \vec{\bf q}_\perp = \vec{P}_\perp$, we get
\bea
\label{LFD2ab}
\Sigma_{\rm LFD}  & \equiv & \Sigma_{\rm LFD}^a =  \frac{1}{2P^+P^--\vec{\bf P}_{\perp}^2-m^2} \nonumber \\
                  & = & \frac{1}{s-m^2}.
\eea 
This result is frame independent and thus valid even in the IMF limit, or $P_z \rightarrow \infty$. 

Finally, let us consider the case of an arbitrary interpolating angle in the range of $0 < \delta < \frac{\pi}{4}$. 
The contribution of diagram of Fig.\ref{fig:2}a is given by
\be
\Sigma^a_{\delta} = \frac{1}{2\omega_q} \left(\frac {1}{P_{\wh+}+\frac{\mathbb{S}q_{\wh-}-\omega_q}{\mathbb{C}}} \right )\, ,\ee
where $ \omega_q = \sqrt{q_{\wh-}^2+\mathbb{C}\left(\vec{\bf q}_\perp^2+m^2 \right)}$.
Since $P_{\wh-} = q_{\wh-}$ and $\vec{\bf P}_{\wh{\perp}} = \vec{\bf q}_{\wh{\perp}}$, we can rewrite these expressions as
\be
\Sigma^a_{\delta} = \frac{1}{2\omega_q} \left(\frac {\mathbb{C}}{\mathbb{C}P_{\wh+}+\mathbb{S}P_{\wh-}-\omega_q}\right)\,;\quad \omega_q = \sqrt{P_{\wh-}^2+\mathbb{C}\left(\vec{\bf P}_\perp^2+m^2 \right)}.
\ee
Using Eq.(\ref{definitions}), we can further reduce the time-ordered amplitude of Fig.\ref{fig:2}a as 
\be
\Sigma^a_{\delta} = \frac{\mathbb{C}}{2\omega_q P^{\wh+}-2\omega_q^2}.
\ee
Since $P^{\wh+} = P^0 \cos\delta + P^3 \sin\delta$ from Eq. (\ref{interpolangle}), we can express $P^0$ in terms of $P^3$  using the dispersion relation $P^2 = s$ as
\bea
P^0 & = & P^3 + \frac{\vec{\bf P}_{\perp}^2+s}{2P^3} + {\mathcal O}\left(\frac{1}{(P^3)^3}\right) \nonumber \\
    & = & P_z + \frac{\vec{\bf P}_{\perp}^2+s}{2P_z} + {\mathcal O}\left(\frac{1}{P_z^3}\right) \, .
\eea
Thus, we get 
\be
\label{P+}
P^{\wh+} = P_z(\sin\delta + \cos\delta) + \frac{\vec{\bf P}_{\perp}^2+s}{2P_z}\cos\delta + {\mathcal O}\left(\frac{1}{P_z^3}\right) \, ,
\ee
and similarly
\be
\label{P-}
P_{\wh-} = P_z(\sin\delta + \cos\delta) + \frac{\vec{\bf P}_{\perp}^2+s}{2P_z}\sin\delta + {\mathcal O}\left(\frac{1}{P_z^3}\right) \, .
\ee
The result given by Eq. (\ref{P-}) is used to evaluate $\omega_q^2$:
\be
\label{omega2}
\omega_q^2 = P_z^2(\sin\delta+\cos\delta)^2 + \left(\vec{\bf P}_{\perp}^2+s\right)\sin\delta(\sin\delta+\cos\delta)+ \mathbb{C}\left(\vec{\bf P}_{\perp}^2+m^2\right) + {\mathcal O}\left(\frac{1}{P_z^2}\right) \, ,
\ee
which leads to  
\be
\omega_q = P_z(\sin\delta+\cos\delta) + \frac{\left(\vec{\bf P}_{\perp}^2+s\right)}{2P_z}\sin\delta + \frac{\left(\vec{\bf P}_{\perp}^2+m^2\right)}{2P_z}(\cos\delta-\sin\delta) + {\mathcal O}\left(\frac{1}{P_z^2}\right),
\ee
where we used the identity 
$$\mathbb{C} \equiv \cos2\delta = \cos^2\delta-\sin^2\delta = (\cos\delta+\sin\delta)(\cos\delta-\sin\delta) \, .
$$
Putting all the ingredients to calculate the denominator, we obtain
\bea
2\omega_q P^{\wh+}-2\omega_q^2 & = & 2P_z^2 (\sin\delta+\cos\delta)^2+\left(\vec{\bf P}_{\perp}^2+s\right)(\sin\delta+\cos\delta)^2+\mathbb{C}\left(\vec{\bf P}_{\perp}^2+m^2\right) \nonumber \\
                               & - & 2P_z^2 (\sin\delta+\cos\delta)^2-2\left(\vec{\bf P}_{\perp}^2+s\right)(\sin^2\delta+\sin\delta\cos\delta)-2\mathbb{C}\left(\vec{\bf P}_{\perp}^2+m^2\right)\nonumber \\
                               & + & {\mathcal O}\left(\frac{1}{P_z^2}\right) \nonumber \\
                               & = & \mathbb{C}(s-m^2)+{\mathcal O}\left(\frac{1}{P_z^2}\right).
                               \eea
This leads to 
\be
\Sigma^a_{\delta}  =  \frac{\mathbb{C}}{2\omega_q P^{\wh+}-2\omega_q^2} \xrightarrow{P_z \rightarrow \infty}  \frac{1}{s-m^2}.
\ee
For the diagram of Fig.\ref{fig:2}b, since 
\bea
2\omega_q P^{\wh+}+2\omega_q^2 & = & 4P_z^2 (\sin\delta+\cos\delta)^2+\left(\vec{\bf P}_{\perp}^2+s\right)(3\sin^2\delta+\cos^2\delta^2+4\sin\delta \cos\delta) \nonumber \\
                               & + & 3\mathbb{C}(\vec{\bf P}_{\perp}^2+m^2)+{\mathcal O}\left(\frac{1}{P_z^2}\right) \, ,
                                \eea
we get 
\be
\Sigma^b_{\delta}  =  \frac{\mathbb{C}}{2\omega_q P^{\wh+}+2\omega_q^2} \xrightarrow{P_z \rightarrow \infty}  0.
\ee

\end{document}